\documentclass[review]{elsarticle}
\usepackage{etoolbox}
\usepackage[a4paper, margin=1in]{geometry}
\makeatletter
\patchcmd{\ps@pprintTitle}{\footnotesize\itshape
       Preprint submitted to \ifx\@journal\@empty Elsevier
       \else\@journal\fi\hfill\today}{\footnotesize\itshape
       Published at European Journal of Operational Research \ifx\@journal\@empty 
       \else\@journal\fi\hfill EOR16143/EJOR-D-18-00153 }{}{}
\makeatother
\usepackage{lineno,hyperref}
\usepackage{adjustbox}
\usepackage{algorithm,algorithmic}
\usepackage{multirow}
\usepackage{pdflscape}
\usepackage{amsmath}
\usepackage{amsfonts}
\usepackage{caption}
\usepackage{amssymb}
\usepackage{subcaption}
\usepackage{todonotes}
\usepackage{siunitx}
\usepackage{rotating}
\usepackage{graphicx}
\usepackage{calrsfs}
\usepackage{float}
\usepackage{chngpage}
\usepackage{natbib}
\usepackage[flushleft]{threeparttable}
\usepackage{booktabs} 
\usepackage{pbox}
\usepackage{anyfontsize}
\usepackage{comment}
\usepackage{xcolor}
\usepackage{makecell}

\usepackage{adjustbox}
\usepackage{array}
\newcolumntype{x}[1]{>{\centering\let\newline\\\arraybackslash\hspace{0pt}}p{#1}}

\makeatletter
\newcommand{\ALOOP}[1]{\ALC@it\algorithmicloop\ #1
  \begin{ALC@loop}}
\newcommand{\ENDALOOP}{\end{ALC@loop}\ALC@it\algorithmicendloop}

\makeatother
\modulolinenumbers[5]
\usepackage{hyperref}



\journal{}

\begin{document}

\begin{frontmatter}

\title{Can Deep Learning Predict Risky Retail Investors? A Case Study in Financial Risk Behavior Forecasting}


\author[ucl]{Y. Yang$^{\dag,}$}
\ead{yaodong.yang@outlook.com}
\author[hub]{A. Kim$^{\dag,}$}
\ead{alisa.k@protonmail.com}
\author[hub]{S. Lessmann$^{\dag,}$}
\ead{stefan.lessmann@hu-berlin.de}
\author[soton]{T. Ma$^{\dag,}$}
\ead{tiejun.ma@soton.ac.uk}
\author[soton]{M.-C. Sung$^{\dag,}$\corref{mycorrespondingauthor}}
\ead{M.SUNG@soton.ac.uk [Corresponding author]}
\author[soton]{J.E.V. Johnson$^{\dag,}$}
\ead{J.E.Johnson@soton.ac.uk}

\cortext[mycorrespondingauthor]{$^\dag$Authors contribute equally. Published at European Journal of Operational Research (EOR16143/EJOR-D-18-00153).}
\address[ucl]{Department of Computer Science, University College London}
\address[hub]{School of Business and Economics, Humboldt-University of Berlin}
\address[soton]{Southampton Business School, University of Southampton}

\begin{abstract}
The paper examines the potential of deep learning to support decisions in financial risk management. We
develop a deep learning model for predicting whether individual spread traders secure profits from future trades. This task embodies typical modeling challenges faced in risk and behavior forecasting. Conventional machine learning requires data that is representative of the feature-target relationship and relies on the often costly development, maintenance, and revision of handcrafted features. Consequently, modeling highly variable, heterogeneous patterns such as trader behavior is challenging. Deep learning promises a remedy. Learning hierarchical distributed representations of the data in an automatic manner (e.g. risk taking behavior), it uncovers generative features that determine the target (e.g., trader's profitability), avoids manual feature engineering, and is more robust toward change (e.g. dynamic market conditions). The results of employing a deep network for operational risk forecasting confirm the feature learning capability of deep learning, provide guidance on designing a suitable network architecture and demonstrate the superiority of deep learning over machine learning and rule-based benchmarks. 

\end{abstract}

\begin{keyword}
risk management \sep retail finance \sep forecasting \sep deep learning
\end{keyword}

\end{frontmatter}

\section{Introduction}
The paper applies recently developed deep learning (DL) methods to forecast the  behavior of retail investors in the spread-trading market. Market makers depend on accurate forecasts of traders' future success to manage financial risks. Through developing a DL-based forecasting model and confirming the profitability of a model-based hedging strategy, we provide evidence that characteristic features of DL generalize to the structured data sets commonly employed in retail finance and decision support. 

Deep neural networks (DNN) operate in a stage-wise manner. Each layer receives an input from previous layers, learns a high-level representation of the input, and passes the representation (i.e., output) to a subsequent layer. Applications in face recognition exemplify this approach. To detect faces in an image, first DNN layers learn low-level concepts such as lines and borders from raw pixels. Deeper layers generalize lower layer outputs to more complex concepts such as squares and triangles that eventually form a face \citep{naturereview}. An analogous example in decision support could be corporate credit risk modeling. Bankruptcy prediction models estimate default probabilities on the basis of ratios of accounting variables (e.g., total assets/total liabilities) \citep{ej_Geng}. In a DL framework, such ratios represent a low level representation. Using balance sheet figures as input, lower layers in a DNN can relate statement variables and calculate informative ratios in a data-driven manner. A higher level representation of the data could then include the trend in a financial ratio or inter-dependencies between ratio variables. The specific representation is calculated autonomously. A hierarchical composition of representations of different complexities enable a DNN to learn abstract concepts such as that of a delinquent borrower. Representation learning also enhances the ability of a model to extract patterns that are not well represented in the training data, which is a problem for machine learning (ML) models \citep{bengio2009learning}. DL methods have delivered excellent results in applications such as computer vision, language processing, and others \citep{Schmidhuber_survey}. This success has established the effectiveness of DL-based feature learning in applications that rely on unstructured data \citep{dl_survey_architecture}. 

Applications of conventional ML are manifold. Marketing models support all stages of the customer life cycle including response modeling, cross-/up-selling \citep{ej_crosssel}, and churn prediction \citep{ej_verbeke}. Financial institutions use ML to anticipate financial market developments \citep{ejor_Oztekin_16}, predict the solvency of corporate borrowers \citep{ej_jardin}, or inform credit approval decisions \citep{cs_bench}. 
Such applications rely on structured data such as past customer transactions, price developments, or loan repayments. It is not obvious that the success of DL in unstructured data processing generalizes to decision support applications where structured data prevails. 
Therefore, the objectives of the paper are to examine the effectiveness of DL in decision support, to test whether its feature learning ability generalizes to the structured data sets typically encountered in this field, and to offer guidance on how to setup a DL-based decision support model.   

We pursue our objectives in a financial risk management context. Using data from the spread-trading market, we predict the profitability of individual traders. The modeling goal is to identify traders that pose a high risk to the market maker, and recommend a hedging policy that maximizes the marker maker's profits. Beyond the utility of such a policy for a spread-trading company, the trader risk prediction task represents challenges that are commonly encountered in ML-based decision support. 

A first challenge is class imbalance. Adverse events such as borrower default represent minorities in their populations, and this impedes ML \citep{ej_verbeke}. A second challenge called concept drift arises in dynamic environments. ML models learn a functional relationship between subject characteristics (e.g., previous trades of a client) and a target (e.g., trader profitability) from past data. Changes in the environment render this relationship more volatile and harder to infer. The curse-of-dimensionality is another modeling challenge. Corporate data warehouses provide a huge amount of information about modeling subjects (e.g., traders) and it is difficult to learn generalizable patterns in the presence of a large number of features \citep{stat_learning}. Finally, the success of ML depends on the availability of informative features. Feature engineering is carried out manually by domain experts. Given high labor costs, a shortage of skilled analysts 
and the need to revise hand-crafted features in response to external changes (e.g., in trader behavior), manual feature engineering decreases the efficiency of ML and becomes an impediment to ML-based decision support. 

The common denominator of the modeling challenges is that they reduce the representativeness of the training data. Being a data-driven approach, ML suffers from reduced representativeness, which suggests that the challenges diminish the effectiveness and efficiency of data-driven ML models. Considering our application setting as an example, the representation learning ability of DL could help to identify a more generic representation of the trading profile of high-risk traders than that embodied in hand-crafted features. More generality in the inferred feature-target-relationship would offer higher robustness toward external variations in trading behavior; for example, variations introduced by changes in  business cycle, market conditions, company operations, etc. Replacing the need for costly manual feature engineering would also raise the efficiency of model-based decision support. 

Examining the degree to which DL remedies common modeling challenges in decision support, the paper makes the following contributions. First, it is one of the first studies to examine the effectiveness of DL in conjunction with structured, individual-level behavioral customer data. Predicting individual trader's risk taking behavior, we focus on retail finance, which is a pivotal application area for operations research \citep{ej_cs_survey_Crook07} 
that, to our knowledge, no previous DL study has considered. Empirical results provide evidence that DL predicts substantially more accurately than ML methods. Second, we demonstrate the ability of DL to learn informative features from operational data in an automatic manner. Prior research has confirmed this ability for unstructured data \citep{naturereview}. We expand previous results to transactional and behavioral customer data. This finding is managerially meaningful because many enterprises employ structured data for decision support. Third, the paper contributes to financial risk taking forecasting practice in that it proposes a DNN-based approach to effectively manage risk and inform hedging decisions in a speculative financial market. 

The DL methodology that we employ in the paper is not new. 
However, DL and its constituent concepts such as distributed representations are rarely explained in the language of business functions. Business users can benefit from an understanding of DL concepts to enable them to engage with data scientists and consultants on an informed basis. A better understanding might also lead to more appreciation of formal, mathematical models and help to overcome organizational inertia, which is a well-known impediment to fact-based decision support \citep{jm_lilien11, Chen_BI2.0}.
Against this background, a final contribution of the paper is that it increases awareness of DL in business through evidencing its potential and providing a concrete recipe for how to set up, train, and implement a DNN-based decision support approach. To achieve this, we elaborate on the methodological underpinnings of DL and the decision model we devise for trader risk classification.  We note that a similar objective has independently been pursued in a recent, related paper by Kraus et al. \citep{ej_dl_feuerriegel}. \color{black}


\section{Related Work}
The literature on DL is growing at high pace \citep{naturereview,Schmidhuber_survey, dl_survey_architecture}. We focus on DL applications in finance. Table \ref{tab_fin_lit} analyzes corresponding studies along different dimensions related to the forecasting setting, underlying data, and neural network topology. To clarify the selection of papers, we acknowledge that DL has other applications in finance beyond forecasting including index tracking \citep{dl_for_finance} or modeling state dynamics in limit-order-book data \citep{ Sirignano_orderbook}. DL has also been applied to generate financial forecasts from textual data \citep{feuerriegel_dss_fin_dis}. Table \ref{tab_fin_lit} does not include such studies as they do not concentrate on prediction or consider a different source of data. Finally, one may argue that a recurrent neural network (RNN) is a DNN by definition, because recurrent cells exhibit temporal depth. With the rise of DL, gated RNNs such as LSTM (long short-term memory) gained  popularity and are often characterized as DNNs \citep{dl_fin_fischer18}. This is not necessarily true for their predecessors, some of which have been used in finance \citep{ej_huck_09}. Table \ref{tab_fin_lit} analyzes studies that used contemporary gated RNNs and omits those that use earlier types of RNNs.

Table \ref{tab_fin_lit} shows that the majority (roughly 60\%) of previous studies forecast developments in financial markets, such as price movements \cite{rl_trading}, volatility \cite{dl_fin_xiong16} or market crashes \cite{dl_fin_chatzis18}. Applications in risk analytics such as financial distress prediction \citep{ dl_fin_addo18} or credit scoring \citep{dl_mortgage} are also popular. Considering the objectives of forecasting, columns two and three reveal that previous studies have not considered forecasting human behavior, which is the focus of this paper. 

The type of input data represents a second difference between most previous studies and this paper. DNNs that forecast financial market prices typically receive lagged prices as inputs. For example, \citep{rl_trading} and \citep{ dl_fin_fischer18} use minute- and day-level price returns as inputs. By contrast, the risk modeling task we face consists of a dynamic regression problem with different types of predictor variables (see Section \ref{subsec:data}). The \textit{feature} columns in Table \ref{tab_fin_lit} show that few prior studies mix numerical and discrete input variables. 

A core feature of DNNs is the ability to automatically extract predictive features from the input data \citep{montufar2014number}. One objective of this paper is to confirm the feature learning capability in a risk management context. A substantial difference in the type of input data has implications for feature learning. It is not obvious that results observed in a time series setting generalize to a dynamic regression setting with diverse input variables. With respect to risk management, we observe from the column \textit{profit simulation} in Table \ref{tab_fin_lit} that most previous work has not examined the economic implications of a DL-based risk management approach; \citep{dl_mortgage} being an  exception.

In addition to the application setting and input data, a third difference between most previous work and this study concerns the architecture of the DNN. Table \ref{tab_fin_lit} sketches the topology of previous networks in its three rightmost columns. Given our focus on forecasting studies, every network includes a supervised learning mechanism, meaning that weights in the network are trained through minimizing the empirical loss on the training data set \citep{Bengio-et-al-2015-Book}. This is typically implemented by means of a fully-connected output layer. This layer requires only one unit with a linear or softmax activation function to solve regression and classification problems, respectively. Table \ref{tab_fin_lit} shows that purely supervised learning networks prevail in previous work. From this observation, we conclude that more research into networks with supervised and unsupervised layers is desirable.

In total, nine studies consider unsupervised pre-training. The majority implement pre-training using a deep belief network. Long before pre-training was popularized, a seminal study proposed self-organizing maps for unsupervised time series pattern extraction \citep{ Giles2001}. Stacked denoising auto-encoders (SdA), the approach we use for feature learning, have received little attention. 
Evidence of their effectiveness in risk analytics is originally provided in this paper. 

In summary, the contribution of our work to literature emerges through a combination of characteristics concerning the forecasting setting, the data employed, and the way in which we devise and assess the DL-based forecasting model through using state-of-the-art approaches for network training and unsupervised pre-training and evaluation of the profitability of model-based hedging decisions. 

The study closest related to our work is \citep{dl_mortgage}. The authors estimate a DNN from a data set of over 3.5 billion loan-month observations with 272 variables relating to loan characteristics and local economic factors to support portfolio management. To that end, \cite{dl_mortgage} model the transition probabilities of individual loans between states ranging from current over different delayed payment states to delinquency or foreclosure. Our study differs from \cite{dl_mortgage} in terms of the application setting and DL methodology. 

The DL models of \cite{dl_mortgage} consists of feed-forward networks of up to 7 layers (and ensembles thereof). Deep feed-forward networks are a generalization of the three-layer networks widely used in previous work (\cite{ej_ann}). The DNN architecture proposed here is different. It uses multiple layers of different types of units and relies on unsupervised pre-training to extract predictive features. Pre-training elements provide distinctive advantages and have been found effective in financial applications \citep{dl_for_finance}. Consequently, we further advance the methodology of \citep{dl_mortgage}.

Mortgage risk modeling \citep{dl_mortgage} and credit scoring in general, differ substantially from trader risk prediction. A credit product can be considered a put option. The lender has the right to grant credit, but no obligation to do so. Credits may also be secured by collateral and, most importantly, it is possible to hedge risks while still earning money from commissions. However, we consider a spread-trading context where the market maker is \textit{obliged} to accept orders from its clients. These orders are similar to futures contracts with an arbitrary strike date. Unlike in the credit industry where customers are given a credit limit, in the spread trading market, informed traders or insiders can make unlimited profits from the market marker. At the same time, the economics of the spread-trading market require the market maker to hedge risks very selectively because hedging quickly reduces revenues to zero. Thus, our forecasting task is to identify those traders who pose a substantial risk to the market maker. 

\section{Risk Taking and Behavior Forecasting in the Spread-Trading Market}
\label{sec:rm}
Spread trading is becoming increasingly significant. Forty percent of the £1.2 trillion traded annually on the London Stock Exchange is equity derivative related and 25 percent of this (£120 billion) relates to spread trading \citep{brady2006white}. Spread trading often refers to pairs trading of stocks or to trading spreads in the futures market \citep{ej_huck_10}. However, our study focuses on the form of spread trading which relates to retail \textit{contracts for difference (CFD)}. In this market, a retail investor and a market maker enter a contract related to a specified financial instrument (e.g. a share, commodity or an index) and at the end of the contract they exchange the difference between the closing and opening price of that financial instrument. Consequently, investors trade on the direction and magnitude of movements of a financial instrument. For example, a client might place a long order on the S\&P500 with stake size \$10 per point. If the S\&P500 rises by a particular $increment$, the client makes a profit of $\$10*increment$; otherwise s/he loses this amount. The market maker quotes bid and ask prices for marketable instruments. Unlike brokers, who help clients to trade with other investors, market makers buy or sell financial instruments from their own inventory. Provided clients meet a margin requirement, they can open and close positions at any time. The market maker is obliged to accept these orders and faces the risk of adverse selection.

\newgeometry{left=1.5cm,bottom=1.5 cm}
\thispagestyle{empty}
\begin{landscape}
 
\begin{table} 
\caption{Summary of Related Work on DL in Finance}
\label{tab_fin_lit}
\begin{adjustbox}{max width=\linewidth}
\begin{threeparttable}
\begin{tabular} {ccx{1.4cm}x{2.5cm}x{1.2cm}x{3cm}x{1.7cm}cccx{3cm}x{2.5cm}x{1.2cm}x{2cm}x{2cm}x{2.6cm}}
\hline
& Area\tnote{1} & Subject\tnote{2} & Target & Time Series & Time Window & Obser-  vations & \multicolumn{2}{c}{Features\tnote{3}} & Horizon & Study Design & Data part.\tnote{4} & Profit sim. & Supervised Layers\tnote{5} & Unsup. Pretrain.\tnote{5} & Architecture\tnote{6} \\ \hline
\cite{Giles2001} & MM & Exr & Direction & 5 & 9/1973 - 5/1987 & 3,645 & 2 & con & day & rolling window & 100 / 30 &  & RNN & SOM & x-7-2-o \\ 
\cite{dnn_fin_shen15} &  & Exr & Return & 3 & 1976 - 2004 & $<1,000$ & 5 & con & week & temporal split & 0.7 / 0.3 &  & FC & DBM & x-20-20-20-o \\ 
\cite{dl_fin_xiong16} &  & Ind & Volatility & 1 & 10/2004 - 7/2015 & 2,682 & 25 (t3) & con & day & temporal split & 0.7 / 0.3 &  & LSTM &  & x-1-o \\ 
\cite{Dixon_fin_market_pred}&  & Fut & Direction & 43 & 3/1991 - 9/2014 & 50 & 9895 & con & 5 min & rolling window & 25000 / 12500 & yes & FC &  & x-1000-100-100-100-o \\ 
\cite{rl_trading} &  & Fut & Return & 4 & 1/2014 - 9/2015 & 100 & 150 & con & minute & rolling window & 15000 / 5000 & yes & RLRNN &  & x-128-128-128-20-o \\ 
\cite{dl_fin_bao17}&  & Ind & Return & 6 & 7/2008 - 9/2016 & 2,079 & 19
(t4) & con & day & rolling window & 2y / 1q / 1q\tnote{7} & yes & LSTM & SdA & x-10-10-10-10-10-1-1-1-1-1-o \\ 
\cite{krauss2017deep} &  & Sto & Better S\&P500 & $\sim$500 & 1/1992 - 10/2015 & 380 & 31 & con & day & rolling window & 750 / 250 & yes & FC &  & x-31-10-5-o \\ 
\cite{dl_crude_oil}&  & Co & {WTI} crude oil spot price & 1 & 1/1986 - 5/2016 & 365 & 200 & con & month & temporal split & 0.80 / 0.20 &  & FC & SdA & x-100-10-o \\
\cite{dl_fin_fischer18}&  & Sto & Better S\&P500 & $\sim$500 & 1/1992 - 10/2015 & 380 & 1 (t240) & con & day & rolling window & 750 / 250 & yes & LSTM &  & x-25-o \\ 
\cite{dl_fin_baek18}&  & Ind & Return & 2 & 1/2000 - 7/2017 & 4,3 & 6 (t20) & con & day & temporal split & 0.45 / 0.55 & yes & LSTM+ LSTM+FC &  & x-(x1-5-3 | x2-4-2)-2-o \\ 
\cite{dl_fin_chatzis18}&  & Ind & Crash & 2 & 1/1996 - 12/2017 & 5,4 & 131 & con & 1 $\mid$ 5 days & temporal split & 0.66 / 0.33 &  & FC &  & x-64-32-8-2-o \\ 
\cite{dl_fin_kim18}&  & Ind & Volatility & 1 & 1/2001 - 1/2017 & 3,963 & 6 (t22) & con & day & temporal split & 0.68 / 0.32 &  & LSTM+FC &  & x-10-4-2-5-o \\ 
\cite{ej_dl_huck} &  & Sto & Better S\&P & 300 & 1/1993 - 5/2015 & 6300 & $\leq 592$ & con & day & rolling window & 504 / 126 & yes & FC & DBN & x-148-74-o \\ 
\hline
\cite{ribeiro11}& RA & Ent & Insolvency & N.A. & 2002 - 2006 & 1,2 & 30 & con & Year & random split & 800 / 400 &  & FC & DBN & x-500-500-1000-o \\ 
\cite{dnn_fin_yeh16}&  & Ent & Insolvency & N.A. & 2001 - 2011 & $\sim83,000$ & 180 & con & Year & rolling window & 04.01.2001 &  & FC & DBN & x-1000-1000-1000-o \\ 
\cite{dl_fin_lee17}&  & Ent & Firm perf. & 22 & 2000 - 2015 & 286 & 15 & con & Year & temporal split & 10. Mrz &  & FC & DBN & x-200-200-200-200-o \\ 
\cite{dl_fin_luo17}&  & Ent & Rating & N.A. & 1/2016 - 12/2016 & 661 & 11 & mix & N.A. & cross-val. & 10 fold &  & FC & DBN & Not specified \\ 
\cite{dl_mortgage}&  & Mor & Default & N.A. & 1/1995 - 6/2014 & $3.5*10^9$ & 272 & mix & month & temporal split & 214 / 19 & yes & FC &  & x-200-140-140-140-140-o \\ 
\cite{dl_fin_addo18}&  & Ent & Insolvency & 286 & 2016 - 2017 & 117,019 & 181 & con & N.A. & random split & 0.6/0.2/ 0.2 &  & FC &  & x-50-50-1 \\
\hline
\cite{dl_fin_cc_fraud}& FD & CC & Fraud & 2 & 5/2015 - 5/2015 & $1.65*10^7$ & 30 (t10) & mix & N.A. & temporal split &0.43/ 0.08/ 0.49&  & LSTM &  & x-100-100-100-o \\ 

\bottomrule \bottomrule 
\end{tabular}
\begin{tablenotes}
            \item[1] {\footnotesize MM: financial market modeling, RA: risk analytics, FD: fraud detection}
            \item[2] {\footnotesize Exr: exchange rate of a pair of currencies, Ind: financial market index, Fut: future contract,  Sto: individual stock, Co: commodity, Ent: enterprise, Mor: mortgage,  CC: credit card.}
            \item[3] {\footnotesize  Number of input features and their type using abbreviations con (continuous feature) and mix (continuous and categorical features). For studies that use LSTM networks we also report the length of a time-lagged input sequence using the notation (tl) where t means time and l is the number of lags. For example, [3] consider 25 features and feed the last three observations (days) of each feature into their LSTM.}
            \item[4] {\footnotesize Partitioning of the data for model training, validation, and testing. Fractional numbers represent percentages with respect to the size of the data set while values greater zero depict absolute numbers of observations. The notation is training set size / validation set size / test set size. Not all studies use separate validation data. Then, the two values given in the column represent training set size / test set size. }
            \item[5] {\footnotesize RNN: recurrent neural network, FC: fully-connected layer, LSTM: Long-short-term-memory, RLRNN: reinforcement learning RNN, SOM: self-organizing-map, SdA: stacked denoising auto-encoders, DBM: deep belief network.}
            \item[6] {\footnotesize Symbols x and o represent the multivariate input and scalar output of the network. Numbers give the size of hidden layers. For studies that use pre-training, hidden layer sizes refer to units of the unsupervised layers (e.g., DBM, SOM, or SdA). Exceptions and special cases for complex topologies exists and we elaborate on these in the discussion of the table.}
            \item[7] {\footnotesize The notation is slightly different from other studies. The authors use a rolling window evaluation to train, validate, and test their models using daily prices from two years, one quarter, and one quarter, respectively.}%
        \end{tablenotes}
\end{threeparttable}
\end{adjustbox}
\end{table}
\end{landscape}
\restoregeometry

Forecasting which traders pose the most risk ( i.e. those who are likely to make the most profit) and deciding which risks to hedge into the main market is crucial for market makers. Informed traders might take advantage of inside information and leave the market maker with positions against a market rally. In theory, the potential loss of the market maker from one trade is unbounded. For example, IG Group, the largest retail financial services provider in UK, recently lost $30$ million GBP due to deficient risk control and inflexible hedging strategies. As a result of similar problems, FXCM, the largest  market maker on the global spot FX market, went bankrupt\footnote{See https://www.forbes.com/sites/steveschaefer/2015/01/16/swiss-bank-stunner-claims-victims-currency-broker-fxcm-bludgeoned/\#7e94f5466de0}. 


The spread between quoted bid and ask prices is the main source of revenue of the market maker. For liquid markets, such as those for the S\&P500 or for the USD/GBP the spread is greater in the spread trading market than in the underlying market. However, for less liquid financial instruments (e.g. the DAX or FTSE100 index) the spread is less than that offered in the underlying market. This later situation is often faced by spread trading firms when they need to place large volume transactions into the underlying market for less liquid financial instruments. If the market maker hedges a trade, they lose the potential profit from the spread whether or not the hedging was necessary. The market maker also faces transaction costs to hedge a position, including commission and the higher spreads in some markets when they seek to hedge large volumes. Therefore, designing a predictive classification model that distinguishes A-book clients (i.e. those who pose most risk to the market maker) from B-book clients (those who pose less risk) is vital. The  market maker will hedge positions from A-book clients to protect against losses and will take the risk of the positions from B-book clients to increase profits. Typically, $90$\% of the total revenues come from B-book clients \cite{pryor2011financial}. 

\begin{figure}[H]
\begin{center}
\includegraphics[height=2.0in]{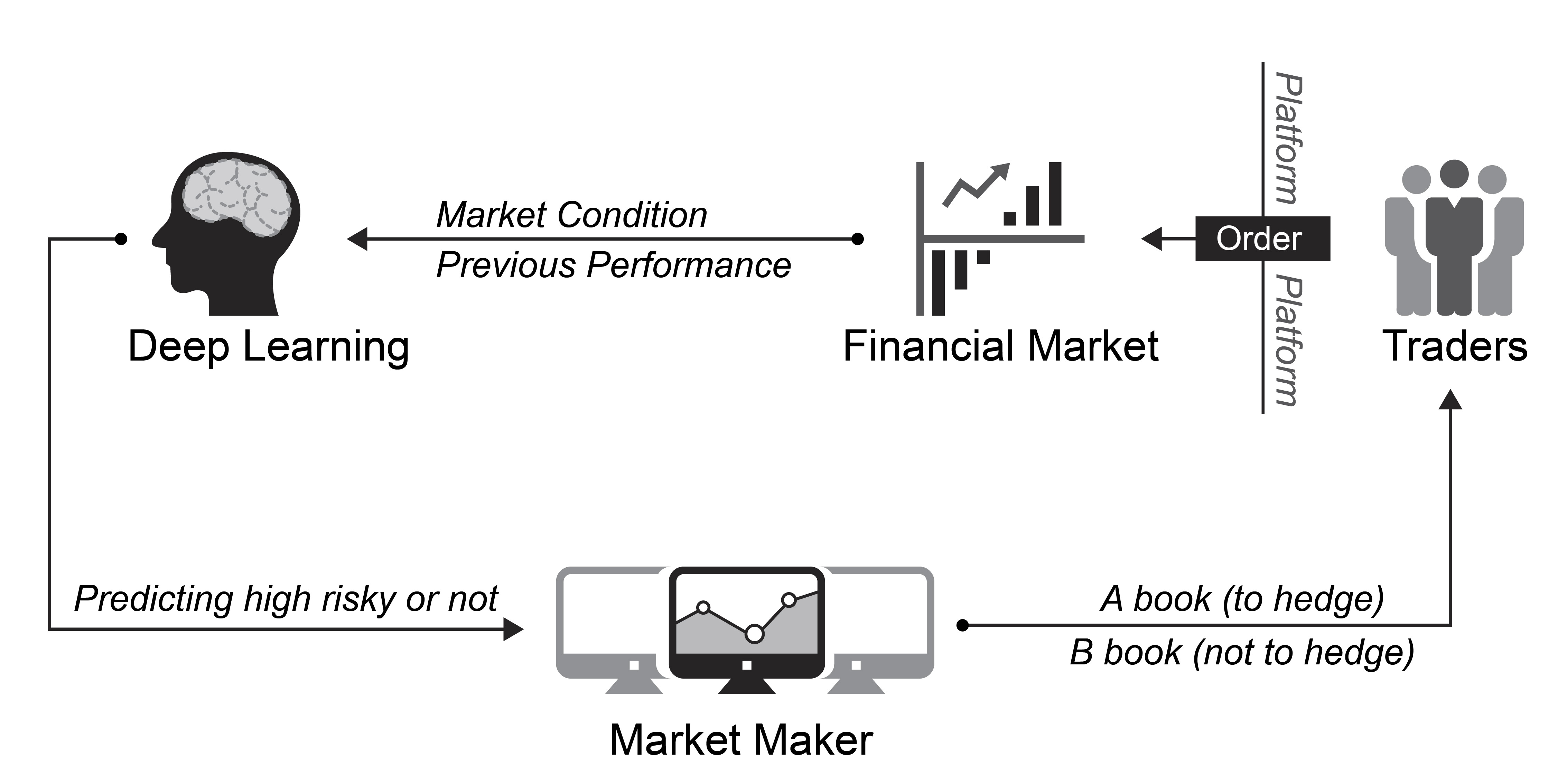}
\caption{Workflow of how hedge strategy works for market makers  }
\label{workflow2}
\end{center}
\end{figure}

We study the decision whether to hedge an incoming trade. This task translates into a classification problem, which we address through developing a DNN to predict high risk (A-book) traders. If the DNN learns patterns from observed trading behavior that facilitate an accurate prediction of a trader's future successes, it can assist the market maker through recommending hedging decisions and enhancing risk management in daily operations. Figure \ref{workflow2} illustrates the DNN-enabled hedging strategy.

\subsection{Trader Classification and Hedging Strategy}
The definition of an A-book client is subjective and depends on the business strategy of the market maker. The company which provided the data prefers to remain anonymous (we refer to them hereafter as STX), but is a large player in the UK spread-trading market. From interviews with their front-desk dealers, who engage in day-to-day risk management, we found that STX at the time of the study, defined a client $i$ to be a high risk trader if s/he secured a return greater than $5\%$ from her previous $20$ trades. The strategy of STX was to hedge the trades of these clients.  

The deployed hedging strategy is dynamic, since STX determines the status of a client (A- or B-book) from the performance of their previous 20 trades. Therefore, client status can change due to a single trade. Accordingly, we frequently observe a situation where STX takes the risk of trade $j$ of client $i$ while hedging against trade $j+k$ of client $i$. In a speculative market, the overall return of a set of past trades can give misleading guidance to the future profitability of a trader. For example, a skilled trader, who follows a consistent strategy, shows high trading discipline, routinely uses and updates stop-loss limits, etc., can regularly lose money due to the randomness of the environment. Similarly, a poor trader, who violates all the above principles, occasionally makes a profit. This suggests that a trader's past performance is not necessarily a reliable signal of their true ability. Consequently, the goal of developing a client classification model is to generate a superior signal for hedging decisions by accounting for all other characteristics available in the data. 

We develop a DNN to learn the latent nature of a trader from past trading data. The target concept, trader ability, is highly variable, corrupted by noise, and difficult to accommodate in a pre-defined, static set of trader characteristics. Therefore, it will be important for the DNN to distill, from transactional data, high-level distributed representations of the target concept, which capture the underlying generative factors that explain variations in trading behavior. In this regard, success in trader classification will evidence the ability of DL to automatically extract informative features.

\subsection{Trader Behavior Prediction and Decision Support}
It is not obvious that representation learning is effective in risk management. Applications such as, credit scoring, churn prediction or trader classification involve the forecasting of human behavior. One would expect the maximal attainable accuracy in a behavior forecasting model to be less than in, e.g., face detection. For example, the prediction target is less clearly defined (e.g., STX used a 5\% threshold but this is subjective) and the feature-target relationship is typically weak. 
Our trader behavior forecasting study aims to clarify the potential of representation learning and DL in decision support.

We argue that the prediction task is representative of a range of modeling challenges in decision support because it exhibits several characteristics that
often occur in corporate applications of data-driven prediction models. More specifically, we face challenges that diminish the representativeness of the training data. First, in response to previous gains and losses and changes in the macro-environment, the behavior of individual traders can be variable, erratic and dynamic. Second, detailed, time-ordered information about individual traders, asset prices and their underlying fundamentals and broader indicators of market sentiment (e.g., economic growth) are readily available, which leads to high dimensionality. Third, the specific way in which variables relate to each other and govern traders' profits  is complex, nonlinear, and likely to evolve over time. Automatic feature extraction, if successful, is a promising way to cope with these challenges. Fourth, the  spread trading setting displays class imbalance in that only a few traders succeed in securing systematic positive returns above $5\%$, while the vast majority of clients lose money. Last, effective risk management requires accurate predictions at the level of an individual trader. Accuracy is a general requirement in predictive decision support.

\section{Methodology}
DL applications in risk analytics are yet sparse. We revisit  principles of DL and detail how we configure the DNN to classify spread traders. The online Appendix elaborates on these concepts and DNN training. %
A recent study on DL for business analytics also provides valuable background \citep{ej_dl_feuerriegel}. 

\subsection{Principles of Deep Learning}
DL aims at learning multiple levels of representations from data. Higher levels represent more abstract concepts. A deep architecture with multiple layers of abstraction and its ability to learn distributed representations provides several advantages over conventional \emph{shallow} ML methods.

\paragraph{The deep architecture}
ML methods learn a functional relationship between variables, which characterize the relationship between an observation and a prediction target. High variability of this function complicates the ML approach and may lead to poor generalization. Sources of high variability include external shocks to the environment in which a decision model operates. Learning theory suggests that to represent a functional relationship, a learning machine with depth $k$ needs exponentially more computational units than a machine with depth $k+1$ \citep{montufar2014number}. The depth of commonly-used machine learning methods is as follows \citep{bengio2009learning}: linear and logistics regression (depth 1); boosting and stacking ensembles: depth of base learner (depth +1, one extra layer for combining the votes from base learners);  decision trees, one-hidden-layer artificial neural networks (ANNs), support vector machines (depth 2); the visual system in the human brain (depth 5-10, \citep{serre2007quantitative}).

The concept of depth explains a large number of empirical findings related to, for example, ANNs or support vector machines outperforming simple regression models or ensemble classifiers outperforming individual learners \citep{cs_bench}. Increased depth allows these methods to implicitly learn an extra level of representation from data \citep{bengio2009learning}. Additional levels facilitate generalization to new combinations of the features, which are less represented in the training data. Enlarged capacity also allows the learning machine to capture more variations in the target function, which discriminates classes accurately. Furthermore, the number of computational units a model can afford is severely restricted by the number of training examples. As a result, when there are variations of interest in the target function, shallow architectures need extreme complexity (large amounts of computational units) to fit the function. Consequently, they need exponentially more training examples than a model with greater depth \citep{bengio2009learning}.


\paragraph{Distributed Representations}
DL methods learn distributed representations from data. An example of a distributed representation is principal component analysis (PCA). PCA re-orients a data set in the direction of the eigenvectors, which are ordered according to their contribution to explained variation. This is a distributed representation where the raw variables collaborate to generate a principle component. In predictive ML, principle components can replace the original variables. The functional relationship to learn is then that between the target variable and the principle components. This can simplify the learning task, increase predictive accuracy, and facilitate feature reduction \citep{is_ulas12}. However, ML methods learn local, non-distributed representations. Using the raw variables in a data set, they partition the input space into mutually exclusive regions. For example, support vector machines infer a decision boundary from the local training examples of adjacent classes that are closest to each other. 

The goal of ML is to classify novel examples, which are not part of the training set. However, the training data may lack representativeness (e.g., because of a change in the environment). Distributed representations are better able to accommodate new observations that the training data does not represent well. Consider our trader classification problem as an example: Traders exhibit different trading styles; they use different strategies, follow different stop-loss rules, etc. Assume traders are split into 5 different clusters, with traders in the same cluster sharing a trading style. Using a non-distributed representation, we need $5$ different features to exclusively represent each cluster, $0=00000,1=01000,..., 4=00001$. A distributed representation requires only $\lceil\log_{2}{5}\rceil=3$ features to model the clusters (as a binary code), $0=000,1=001,..., 4=100$. Using three distributed features, this representation can also accommodate a new type of trader (i.e., using trading strategies that have not been employed in the training sample): $5=101$. This exemplifies an advantage of distributed representations, namely that the number of patterns that the representation can distinguish grows exponentially with the number of features. However, for non-distributed representations, this number grows, at best, linearly.

\subsection{Building the Deep Neural Network} \label{meth_build_dd}
DL methods consist of multiple components with levels of non-linear transformations. A typical instance is a neural network with several hidden layers \citep{krauss2017deep,dl_mortgage}.  
Training a DNN requires solving a non-convex optimization problem, which is difficult because of the vanishing gradient problem \citep{bengio2007greedy}. 
Gradient vanishing prohibits propagating error information from the upper layer back to lower layers in the network, so that connection weights in lower layers cannot be adapted \citep{larochelle2009exploring}. As a result, the optimization will often terminate in poor local minima. Remedies to this problem include unsupervised pre-training, parametric Rectifier Unit (ReLu), Xavier initialization, dropout, and batch normalization. We take advantage of these techniques to develop a trader classification DNN for risk management. Below, we introduce  pre-training and dropout. Interested readers find a similar description of the other concepts in the online Appendix.
    
\subsubsection{Unsupervised Pre-Training}
The goal of pre-training is to find invariant, generative factors (i.e., distributed representations), which explain variations in the data and amplify those variations that are important for subsequent discrimination. Through a sequence of non-linear transformations, pre-training creates layers of inherent feature detectors without requiring data labels. This facilitates a local learning of connection weights. Avoiding a propagation of error information through multiple layers of the network, pre-training helps to overcome the vanishing gradient problem. Stacking multiple layers of progressively more sophisticated feature detectors, the DNN can be initialized to sensible starting values. After discovering a structural relationship in the data, one can then add a supervised learning technique (e.g.,  logistic regression) on top of the pre-trained network and tune parameters using back-propagation. Unsupervised pre-training, where the use of the target label is postponed until the fine-tuning stage, is especially useful in management decision support where class imbalance is a common problem \citep{ej_svm_credit}. 

Two classical implementations of pre-training are deep belief networks (DBN), which are pre-trained by restricted Boltzmann machine \citep{hinton2006fast}, and stacked denoising autoencoders (SdA), which are pre-trained by the autoencoder \citep{bengio2007greedy}. Both strategies minimize an approximation of the log-likelihood of a generative model and, accordingly, typically show similar performance \citep{bengio2009justifying,vincent2008extracting}. This, together with the fact that deep belief networks have already received some attention in the risk analytics literature (see Table \ref{tab_fin_lit}), led us to use the framework of the stacked denoising autoencoder \citep{vincent2008extracting}. 
 
\paragraph{Denoising Autoencoder} The denoising autoencoder (dA) learns a distributed representation (namely the "code") from input samples.
Suppose we have $N$ samples and each sample has $p$ features. Receiving an input $\textbf{x}\in \mathbb{R}^p$, the learning process of a dA includes four steps:

\textbf{{Step $1$}}: \emph{Corruption}. The dA first corrupts the input $\textbf{x}$. By sampling from the Binomial distribution 
$(n=N, p=p_q)$
, (where the corruption rate $q_p$ is a hyper parameter that needs tuning outside the model), the dA randomly corrupts a subset of the observed samples and deliberately introduces noise. For example, if the input features a binary, corruption corresponds to flipping bits.

\textbf{Step $2$}: \emph{Encoder}. The dA deterministically maps the corrupted input  $\widetilde{\textbf{x} }$ into a higher-level representation (the code) $\textbf{y} \in \mathbb{R}^{k}$. The encoding process is conducted via an ordinary one-hidden-layer neural network (the number of hidden units $k$ is a hyper parameter that needs tuning outside the model). With weight matrix $W$, biases $b$, and encoding function $h(\cdot)$, e.g., \emph{sigmoid} function, $\textbf{y}$ is given as:
\begin{equation}
\begin{aligned}
 \textbf{y}=h(W \cdot  \widetilde{\textbf{x}} + b)
\end{aligned}
\end{equation}


\textbf{Step $3$}: \emph{Decoder}. The code \textbf{y} is mapped back by a decoder into the reconstruction \textbf{z} that has the same shape as the input $\textbf{x}$. Given the code $\mathbf{y}$, $\mathbf{z}$ should be regarded as a prediction of $\mathbf{x}$. Such reconstruction represents a \emph{denoising} process; it tries to reconstruct the input from a noisy (corrupted) version of it.
Similar to the encoder, the decoder has the weight matrix $\widetilde{W}$, biases $\widetilde{b}$, and a decoding function $g(\cdot)$. 
The reconstruction $\textbf{z}$ is:
\begin{equation}
\begin{aligned}
\textbf{z}=g(\widetilde{W} \cdot  \textbf{y} + \widetilde{b})
\end{aligned}
\end{equation}

\textbf{Step $4$}: \emph{Training}. Optimizing the parameters of dA ($W$, $b$, $\widetilde{W}$, $\widetilde{b}$) involves minimizing the reconstruction error $L_{(\textbf{x},\textbf{z})}$; achieved by letting the code \textbf{y} learn a distributed representation that captures the main factors of variation in \textbf{x}. Theoretically, if we use the mean squared error ($L_{H(\textbf{x},\textbf{z})} = || \mathbf{x} -\mathbf{z} ||^2$) as the cost function and linear functions as both encoder $h(\cdot)$ and decoder functions $g(\cdot)$, the dA is equivalent to PCA; the $k$ hidden units in code $\textbf{y}$ represent the first $k$ principal components of the data. The choice of cost function depends on the distributional assumptions of input $\textbf{x}$. In this paper, we measure the reconstruction error by the cross entropy loss function, as most of our features are probabilities $\textbf{x} \in [0,1]^p$. In addition, we incorporate an L2 penalty (also called weight decay). This is equivalent to assuming a Gaussian prior over the weights and a common approach to encourage sparsity among weights and improve generalization. The regularization parameter $\lambda$ captures the trade-off between reconstruction error and model complexity. The parameter needs tuning outside the model and offers a way to protect against overfitting. Higher values of $\lambda$ penalize model complexity more heavily and, ceteris paribus, reduce the risk of overfitting. The final cost function is:
\begin{align}
L{(\textbf{x},\textbf{z})}=-\dfrac{1}{N}\sum_{i=1}^{N}\sum_{k=1}^{p}[x_{ik}\log z_{ik}+(1-x_{ik})\log(1-z_{ik})]+ \lambda  \left \| W \right \|_{2} 
\label{cost1}
\end{align}
Several solvers (e.g., stochastic gradient descent) are available to carry out the optimization. 
\begin{equation}
\begin{aligned}
\arg\min_{w,\widetilde{w}, b, \widetilde{b}} L{(\textbf{x},\textbf{z} \hspace{1mm} | \hspace{1mm} \Theta)}
\end{aligned}
\end{equation}

\begin{figure}[htbp]
\begin{center}
\includegraphics[height=2.0in]{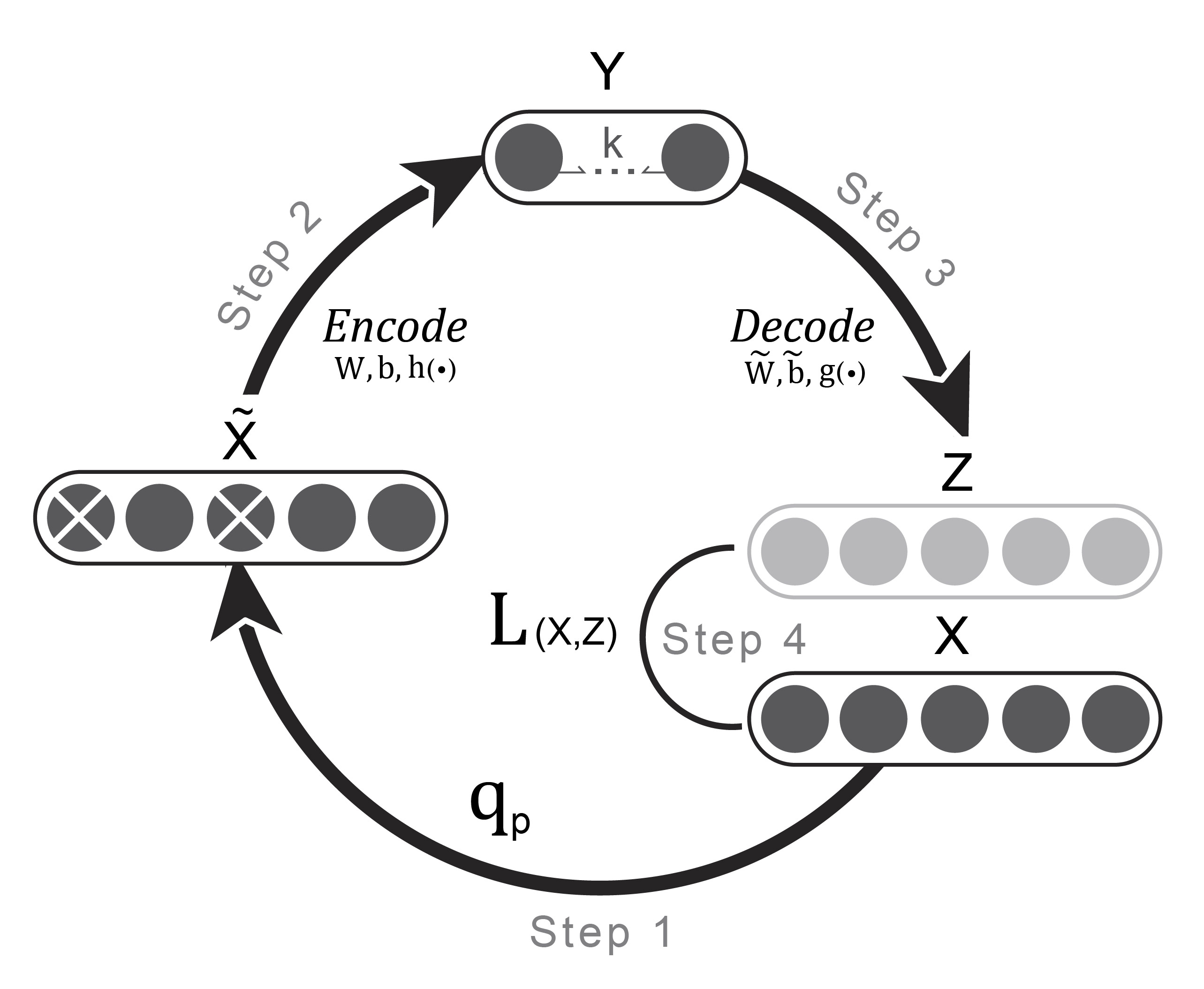}
\caption{Architecture of denoising auto-encoder.  } 
\label{da}
\end{center}
\end{figure}

\textbf{Step $5$}: \emph{Stacking}.
Once a dA has been trained, one can stack another dA on top. Layers are organized in a feed-forward manner. The second dA takes the encoded output of the previous dA (the code \textbf{y}) as its new inputs \textbf{x}. Each layer of dA is trained locally, finding its own optimal weights regardless of the other layers. Iteratively, a number of dAs can be stacked upon each other to construct a stacked denoising autoencoder (SdA). The encoding weights of each dA can then be treated as initializations for the network in the next step. Figure \ref{da} illustrates the working flow of dA. 

\subsubsection{Supervised fine-tuning}
The SdA can be trained in a feed-forward, layer-wise manner. To employ the network for prediction, network training continues with supervised fine-tuning that teaches the DNN which types of trading behaviors (in the form of distributed representations) identify A-book clients. To that end, we add a \emph{softmax} regression on top of the SdA. This way, we solve a supervised learning problem using the distributed representation of the raw input as features (which the SdA output embodies), and a binary indicator variable as target, which indicates whether a trade should be hedged. 
Formally, with parameter weight $W$ and bias $b$, the probability that a trade $\textbf{x}$ belongs to class $i$ is:

\begin{align}
 P(Y=i|x, W,b) = \text{softmax}_i(W x + b) 
 = \frac {e^{W_i x + b_i}} {\sum_j e^{W_j x + b_j}}
\label{softmax}
\end{align}
 We employ the negative log-likelihood as cost function in supervised fine-tuning. Suppose $y^{(i)}$ is the true class for the input $x^{(i)}$, the cost function then states:
\begin{align}
L {(W,b, \textbf{x})} &= 
  -\sum_{i=1}^{N} \log(P(Y=y^{(i)}|x^{(i)}, W,b))
\label{cost2}
\end{align}

\subsubsection{Protecting Against Overfitting Using Dropout Regularization}
\begin{figure}[htbp]
\begin{center}
\includegraphics[height=2.0in]{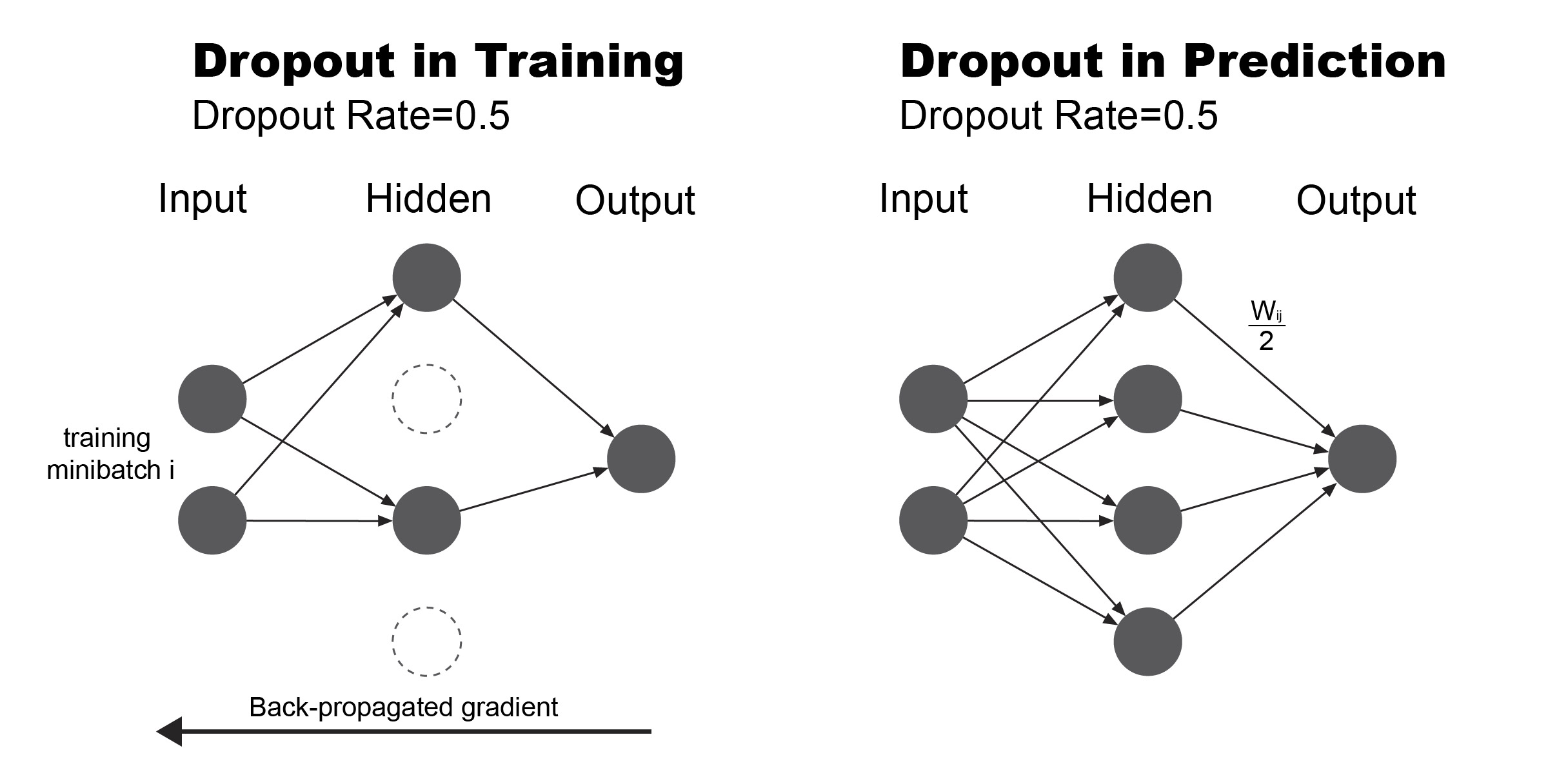}
\caption{Principle of dropout in training and predicting.} \label{dropout}
\end{center}
\end{figure}

DNN are vulnerable to overfitting. To prohibit the model emphasizing idiosyncratic patterns of the training data, our DNN includes a dropout layer behind each hidden layer. 
Figure \ref{dropout} depicts the concept of dropout. During  training, hidden layer neurons and their corresponding connection weights are removed from the DNN. This is done for each batch of training samples in an iteration. The gradients contributed by that batch of samples also bypass the dropped-out neurons during back-propagation (see the online Appendix for a detailed explanation of DNN training). The probability of a hidden neuron dropping out follows a Bernoulli distribution with a given dropout rate. 

A DNN trained with dropout mimics the behavior of an ensemble model. When calculating predictions, the DNN considers all hidden layer neurons but multiplies the connecting weights of each hidden neuron by the expectation of the Bernoulli distribution. This way, although training a single DNN with $N$ hidden neurons, the prediction of the DNN implicitly integrates predictions of $2^N$ candidate networks with different combinations of hidden neurons. More formally, dropout simulates a geometric model averaging process; each possible combination of hidden neurons is considered, which is the extreme case of bagging. Model combination is known to increase predictive accuracy \citep{ej_verbeke,cs_bench}. 

Dropout also acts as a regularizer. It removes random weights from training, which prevents hidden neurons from co-adapting to each other. Moreover, model averaging reduces variance, which, via the bias-variance decomposition, reduces forecast error. Controlling the complexity of a DNN, dropout helps to protect against overfitting \citep{srivastava2014dropout}.
  
Recall that we augment dropout through using an $L2-$penalty during SdA training to increase the robustness of the DNN toward overfitting. For the same reason, we make use of early-stopping.

\subsubsection{Network Training and Configuration}
\label{sec:dnn:tune}
Our DNN involves unsupervised pre-training using SdA. We tune weights in a layer-wiser manner and then fine-tune the DNN as a whole in a supervised way, with each hidden layer followed by a dropout layer. In addition, we use several other DL concepts to protect against overfitting and simplify the network training process including Xavier's initialization, batch normalization layers and using ReLU as the activation function. Previous work on DL has elaborated on these concepts and established their utility \citep{Bengio-et-al-2015-Book}, and we detail them in the online Appendix. %
To facilitate replication, Algorithm 1 in the online Appendix provides a complete description of the employed DNN using pseudo-code. The architecture of the DNN is also sketched in Figure \ref{sda} (below), together with other parts of our methodology. \color{black} 

The parameters to determine in the pre-training stage are the weight matrix and bias in each dA (both the encoder and the decoder). The parameters in the supervised fine-tuning stage are the weight matrix and bias in each encoder of SdA and in the \emph{softmax} regression. We use stochastic gradient descent with momentum and a decaying learning rate for DNN training. The online Appendix provides an  explanation of these concepts and motivates our choices. In particular, Algorithm $1$ in the online Appendix provides a fully-comprehensive description of network training using pseudo-code. Section 2 of the online Appendix also elaborates on our approach to decide on DNN hyper-parameters such as the number of hidden layers in SdA, and how we tune these using random search \citep{bengio2012practical}.  

The techniques we employ are available in DL software packages, which facilitate defining the topology of a DNN, provide routines for numerical optimization to train the DNN, and offer the functionality to apply a trained model for forecasting. 
We use the Python library \emph{Theano}, which is a GPU-based library for scalable computing. The GPUs used for experiments were \emph{Nvidia Tesla K20m} with $2496$ cores and $5$GB GDDR5 high bandwidth memory each. We observe this infrastructure to provide a $10$-$15$ times improvement in speed over training a DNN using traditional CPUs for DNN training (which equates to reducing run-times from more than a week to $1$-$2$ days). In appraising these figures it is important to note that i) large run-times result from the size of the data set, and that ii) training complex ML models may be  as costly. For example, depending on the specific configuration, training a random forest (RF) classifier on the spread trading data can easily require more than 3 days.

\section{Experimental Design}
\label{sec:exp_design}
\begin{figure}[htbp]
\begin{center}
\includegraphics[height=3.8in,width=0.8\textwidth]{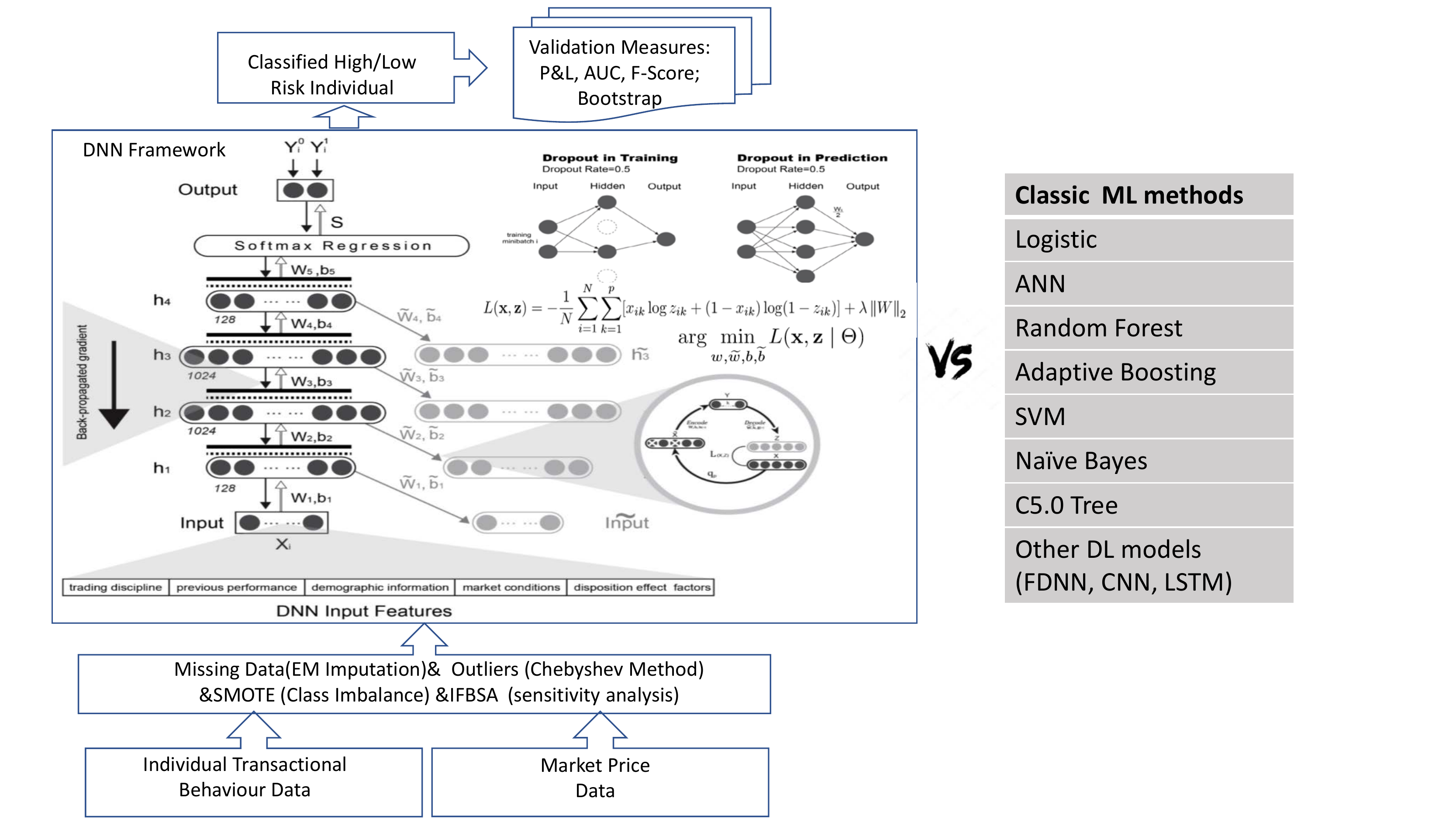}
\caption{Topology of the DNN using sDA with $4$ hidden layers with $128, 1024, 1024, 128$ hidden units each. The output layer predicts class probabilities based on the output of the last dropout layer using the softmax function. DNN predictions are compared to ML benchmarks for assessing predictive performance.} \label{sda}
\end{center}
\end{figure}

The section describes the spread-trading data set, elaborates on the definition of A-book clients, and introduces model evaluation criteria and benchmark classifiers.  Figure \ref{sda} summarizes the empirical design together with core parts of our methodology and the proposed DNN. \color{black}

\subsection{Dataset and Target Label Definition} \label{subsec:data}
STX  provided $11$ years of real-life trading data for the period November 2003 to July 2014. Overall, the data includes the trades of $25,000$ active traders (over $30$ million trades across $6064$ different financial instruments). To prepare the data for analysis, we replaced missing values using EM imputation 
and \emph{Chebyshev}'s method for outlier treatment \citep{stat_learning}.

Supervised learning requires a labeled data set $D =\left\{y_j,\emph{x}_j\right\}_{j=1 \ldots n}$,
where $\emph{x}_j$ is a vector of features that characterize trade $j$, $n=30$ million is the total number of trades in the data, and $y_i$ denotes the target variable. However, data characterizing an individual trade is limited. Relating trades to their corresponding traders facilitates enriching the set of features by using information from previous trades $j-k$ to decide on trade $j$.

The decision task of STX is whether to hedge trade $j$. Therefore, we consider a binary target: 
\begin{eqnarray} \label{def}
y_{ij} & = &\begin{cases}
 & +1 \rightarrow \text{hedge}  ~~,~~\text{iif}~\text{Return}_{i} \geq 5\%\\ 
 & -1  ~~~~~~~~~~~~~~,~~ \text{otherwise}
\end{cases}\\
\text{with}\\
\text{Return}_{i} &= & \dfrac{\sum_{20< j\leq 100}\text{P\&L}_{i,j}}{\sum_{20< j\leq 100}\text{Margin}_{i,j}} \notag 
\end{eqnarray}

where $i,j$ index trader $i$ and trade $j$, respectively, \emph{P}\&\emph{L} is the profit and loss of trade $j$, and \emph{Margin} is the amount of money required by the market maker in order to place the order, which normally equals the stake size times the margin requirement. To label trade $j$, we determine the status of trader $i$ at the time of issuing that trade. We define trader $i$ to be an A-book client if s/he secures a return above 5\% from her next hundred trades subsequent to $j$. Recall that the $5\%$ threshold mimics the current policy of STX. We also sustain the STX approach to hedge all trades from A-book clients. However, our method to define the client status and label their trades is forward looking whereas STX considers the past profits of trader $i$. Our target label definition is also dynamic in that the trader status can change with every trade. According to that definition, $6.43\%$ of the trades in the data set come from A-book clients and should be hedged.

Of course, at the time when STX receives trade $j$, the future profits of trader $i$ are unknown. Therefore, we develop a prediction model to forecast $y_{ij}$ from the information the company can observe at the time when trade $j$ is made. The feature vector $\emph{x}_{ij}$ includes demographic information of the client making trade $j$ and information concerning the client's trading behavior for the 20 trades prior to trade $j$. The decision to consider the past 20 trades is based on the hedging policy of STX, which uses a rolling window of the 20 trades prior to trade $j$ to decide on the status of the client. 

\subsection{Trader Characteristics and Feature Creation}
\label{sec:feature_creation}
We create variables for trader classification based on interviews with experienced members of the dealing desk of STX. A first round of interviews was aimed at identifying risk factors that domain experts deem indicative of good/bad traders. Based on corresponding results, we developed a semi-structured survey that was presented to seven members of the dealing desk in a second round of interviews. The survey asked participants to evaluate behavioral traits, which emerged from the first round, on a Likert Scale from 1-7, where values of 1 and 7 represent a strong indication for bad and good trading behavior, respectively. After completing the survey, we asked participants to suggest strategies they would apply if trading the FTSE100 index and a single stock from the FTSE100, respectively. This was to gather ideas for novel factors not yet covered in the survey. 
The results of the interviews guided the feature engineering. A non-disclosure agreement with STX prohibits formally defining all features.
However, the following description provides a comprehensive overview of the type of features and how they have been created. The features reflect the specific situation of STX. Risk analysts may find the following description useful to inform feature engineering in related applications. However, since our study focuses on the application of a DL methodology, it does not warrant claims related to generalizability of features. In general, features split into five  groups. The first group comprises trader \emph{demographics} such as age, country of origin, post code, employment status and salary group. Features of this group are nominal and enter ML models in the form a dummy codes. 
STX employs a range of socio- and micro-geographic data to cluster post codes. They follow a similar logic to cluster countries.
\footnote{ STX has not revealed details of their cluster mechanisms to us. However, they assured us that the clustering does not employ any information of trader profits, which might otherwise introduce a look-ahead bias through leaking information from the  prediction target to the features.}

Features of the second group capture the \emph{past performance} of traders. We use aggregations such as the mean and standard deviation to calculate corresponding  features over a rolling window of 20 previous trades relative to the focal trade. The choice of a window size of 20 follows STX's  hedging policy at the time of the study. 
In addition to profitability, we compute a set of related performance indicators such as the average win rate, average number of points in profit, whether a client has been in profit, etc. We also consider the risk adjusted return (i.e., Sharpe ratio \cite{DOWD2000209} and features related to the number and sizes of past withdrawals and deposits).

A third group of features describes traders' preferences related to \emph{markets} and \emph{channels}. For example, one feature simply counts the number of markets in which a trader invests while another encodes whether traders showed a strong preference for a specific market in their previous 20 trades. Using this information, we create features describing the most popular market cluster in a trader's full history and last 20 trades, respectively. The subgroup of channel preferences includes features that count the number of opening and closing trades made through the STX web front-end and mobile app, respectively, as well as ratios derived from these counts.

Results of the survey identified the \emph{disposition effect} as a relevant factor to detect poor traders. The disposition effect \citep{weber1998disposition} describes the phenomenon that investors tend to quickly sell trades that are in profit but are reluctant to sell trades in loss. Features of the fourth group strive to capture the disposition effect. We determine per trader the average amount and time s/he leaves winning and loosing positions open, and calculate their ratio. We also consider sums instead of averages and window lengths of the previous 20 and all previous trades. 

Another discriminating factor that emerged from the interviews concerns \emph{trading discipline}. Members of the dealing desk pointed out that good traders display a tendency to set manual limits (stop losses and profit levels) and when making profits to move these with the market. The fifth group of features captures signals concerning the consistency of a trader's strategy. The variation index of stake sizes exemplifies corresponding features. We also consider simpler features such as the standard deviation of stake sizes and features that capture the frequency of trades as well as their variation. Other features in this group relate to the tendency of clients to trade within/outside of normal trading hours (e.g., number and share of corresponding trades), which we consider an indicator of traders' professionalism. The degree to which traders partially close trades may also signal expertise and hence traders posing a higher risk. Hence, we create a feature measuring the share of trades that have been closed in the previous 20 trades.   
The previous examples sketch the type of features we employ. Using operations such as varying window sizes, aggregation functions, creating dummy features through comparing a feature to a threshold (e.g., whether any of the last 20 trades has been closed using the mobile app), and considering bi-variate interactions, we obtain a collection of close to $100$ features. An objective of the paper is to test whether the DNN can learn predictive higher-level features automatically. For example, the discussion on feature engineering suggests multicollinearity among features, which feature selection could remedy. However, a sub-goal following from our objective is to test how effectively the DNN automatically discards redundant and irrelevant features. Therefore, we do not perform feature selection.

\subsection{Exploratory Data Analysis and Feature Importance}
\label{sec:eda}
To shed light on how  A-book and B-book clients differ across the features, we report results of an exploratory data analysis. Table \ref{tab:eda} reports descriptive statistics for the ten top-ranked features for A-book and B-book clients, respectively. We select these features according to the Fisher-score \citep{ej_verbeke}. Features with the suffix 20 are calculated over a window of 20 past trades relative to a focal trade. For example, given a trade $j$ (equivalent to one observation in the data set) from a trader $i$, we consider the $j-1, j-2, ... j-20$ trades of trader i and calculate their mean, standard deviation, etc. We use all available trades of a trader if s/he has less than 20 trades. To ensure comparability across features, values in Table \ref{tab:eda} are scaled to the zero-one interval using min/max scaling. 

 Interestingly, the Fisher-score ranks the feature \emph{PassAvgReturn20}, which determines the current hedging policy of STX, as the tenth most important feature. This suggests that some improvement of the current policy might be possible by basing decisions on one of the more important features such as \emph{ProfitxDur20}, which is the top feature in the Fisher-score ranking. \color{black} More generally, Table \ref{tab:eda} reveals that differences between the client groups in the means of variable values are small. This indicates that good and bad traders cluster in the behavioral space spanned by these features and that a classification of traders will be challenging. To support this view, we estimate a logistic regression model on the training set using the features of Table \ref{tab:eda} and observe a McFadden $R^2$ close to zero. Considering standard deviations, Table \ref{tab:eda} suggests that the trading behavior of B-book clients is slightly more volatile compared to A-book clients, which supports findings from the interviews that good traders follow a consistent strategy. Table \ref{tab:eda} also emphasizes the disposition effect as a potentially discriminating factor. Several of the top ten features aim at capturing the disposition effect through contrasting the duration with which traders keep winning versus losing positions. Last, the third and fourth moment of the feature distributions hint at some differences between good and bad traders. However, as shown by the failure of the logistic regression, translating these differences into a classification rule is difficult and may be impossible with a linear model.   To further substantiate this interpretation, the online Appendix provides an analysis of feature importance using the RF classifier \cite{stat_learning}. RF-based importance weights (see Section 3 of the online Appendix) differ notably from Table \ref{tab:eda} in that the ten most important features in terms of the Fisher-score achieve only moderate importance ranks according to RF. This may evidence the existence of nonlinearity in the feature response relationships or feature interactions, both of which the Fisher-score cannot accommodate. We further elaborate this point in Section \ref{sec:benchmark} when re-appraising feature importance using information fusion-based sensitivity analysis \citep{ifbsa_first, dss_Oztekin_13} \color{black}.

\begin{table}[!htb]
\flushleft
\caption{Descriptive Statistics of Top-Ten Features}
\scalebox{0.82}{
\begin{tabular}{p{3cm}ccccccp{5cm}}
\hline
\multirow{2}{*}{\textbf{Feature} \tnote{1} } & \multicolumn{2}{c}{\textbf{Mean}} & \multicolumn{2}{c}{\textbf{Std.Dev.}} & \multicolumn{2}{c}{\textbf{Skew}} & \multirow{2}{*}{\textbf{Description}}                                                                    \\ \cline{2-7}
                                  & \textbf{A-book} & \textbf{B-book} & \textbf{A-book}   & \textbf{B-book}   & \textbf{A-book} & \textbf{B-book}  &                                                                                                          \\ \hline
ProfitxDur20                      & 0.325           & 0.332           & 0.172             & 0.178             & 0.994           & 0.962                   & Interaction of ProfitRate20 and DurationRate20                                                           \\ 
SharpeRatio20                     & 0.443           & 0.446           & 0.081             & 0.085             & 1.097           & 1.131                       & Mean/st.dev. of returns                                          \\ 
ProfitRate20                      & 0.496           & 0.504           & 0.241             & 0.248             & 0.346           & 0.328                      & Average profit rate of client                                              \\ 
WinTradeRate20                    & 0.621           & 0.626           & 0.203             & 0.207             & -0.203          & -0.210                    & Client’s average winning rate                                                   \\ 
AvgOpen                         & 0.534           & 0.539           & 0.218             & 0.228             & -0.345          & -0.311                 & Average of the P\&L among trader's first 20  trades                                           \\ 
DurationRate20                    & 0.319           & 0.322           & 0.119             & 0.121             & -0.148          & -0.161                    & Average time client leaves winning vs losing position open  \\ 
PerFTSE20                         & 0.251           & 0.244           & 0.357             & 0.353             & 1.151           & 1.197                     & Share of trades placed in the FTSE100                                                                    \\ 
DurationRatio20                   & 0.127           & 0.128           & 0.067             & 0.070             & 3.398           & 3.812                     & Mean trade duration (mins) / std.dev. trade duration                            \\ 
AvgShortSales20                   & 0.487           & 0.482           & 0.269             & 0.274             & -0.027          & -0.018                    & Share of short positions                                                 \\ 
PassAvgReturn20                   & 0.502           & 0.503           & 0.052             & 0.057             & -0.295          & 0.065                     & Avg. return up to last 20 trades                                                                  \\ \bottomrule \bottomrule
\label{tab:eda}
\end{tabular}}
\end{table}


\subsection{Data Organization, Evaluation Criteria and Benchmark Classifiers}
\label{exp_design_bench}
We use $n$-fold cross-validation to assess the predictive performance of ML models. 
Repeating model building and assessment $n$ times increases the robustness of results compared to a single partitioning of the data into  training and test set. We consider settings of $n=10$ and $n=5$ in subsequent comparisons. 

The client classification problem exhibits class imbalance and asymmetric error costs. 
 
A false negative (FP) error corresponds to misclassifying an A-book client as a B-book client. A false positive (FP) error describes the opposite case where a B-client is misclassified as high risk trader. The economic implication of a FP error is that the market maker hedges a trade from a B-client. This is less suitable because B-clients lose money on average. These clients' losses represent profits to the market maker but are driven to zero if a trade is hedged. The FN error is more severe because failing to hedge a trade from a high risk trader may leave the market maker with a very large loss. To reflect this cost asymmetry, we evaluate a classification model in terms of the profit or loss (P\&L) that results from hedging trades according to model predictions.  

P\&L accounts for the actual profits/losses per trade. In addition, we average the P\&L across trades from A-clients and B-clients. The resulting means provide an estimate of the average costs of a FN error, $c_FN$, and a FP error, $c_FP$, respectively. We then compute the average, class-dependent misclassification costs of a classifier per trade as: $c_FN\cdot FNR + c_FP\cdot FPR$ to obtain a second monetary performance measure. Grounding on class-specific averages as opposed to trade-level profits/losses, the average misclassification costs (AMC) is more robust toward the specific trades that a model classifies (in)correctly and outliers, while the P\&L reflects the true monetary implications of deploying a classifier to our sample of data. Hence, the two measures complement each other.

Finally, several standard indicators are available to assess the performance of a classification model. To obtain a broad view on model performance, we consider the area under a receiver-operating-characteristics curve (AUC), sensitivity,  specificity, precision, the G-mean, and the F-measure. The selection of performance indicators draws inspiration from prior literature \citep{oztekin_18, ej_f_measure} and covers a range of popular indicators. A detailed discussion of their merits and demerits is beyond the scope of the paper but is available in the literature \citep{stat_learning}. Roughly speaking, the AUC assesses a classifier's ability to discriminate A- and B-book clients. Sensitivity and specificity depict class-level classifier accuracy and thus the foundation of the AMC measure. Precision emphasizes a model's ability to detect high risk traders while G-mean and the F-measure are designed for robustness toward class imbalance. 
\color{black}

To compare the performance of our DNN to benchmarks, we select five ML classifiers as benchmarks, including logistic regression, ANNs, RF, adaptive boosting (Adaboost), and support vector machines (SVM). A comprehensive description of the classifiers is available in, e.g., \citep{stat_learning}. We report the hyper-parameter settings that we consider during model selection in Section 2 of the online Appendix, where we also elaborate on hyper-parameter tuning. 
 
\section{Empirical Results}
The empirical analysis compares the DNN to benchmark classifiers and rule-based hedging strategies, which embody domain knowledge. It also sheds light on the origins of DNN performance.

\subsection{Predictive Accuracy of the DNN and ML-based Benchmark Classifiers}
\label{sec:benchmark}
 
We first present results concerning the predictive performance of different classifiers across different evaluation criteria in Table \ref{tab:cv10}. Bold face highlights the best performing model per evaluation criteria. P\&L and AMC values are measured in GBP and refer to an individual trade. For example, a P\&L value of 121.67, which we observe for the DNN, implies that hedging incoming trades according to DNN recommendations, the market maker would on average earn 121.67 GBP from each trade. A value of 483.84 for AMC, on the other hand, indicates that an oracle model, which \emph{knows} the outcome of each incoming trade and hedges accordingly, would produce 483.84 GBP profit per trade for the market maker. Clearly, the optimal hedging of an oracle model is a theoretical benchmark. All values in Table \ref{tab:cv10} represent averages, which we obtain from 10-fold cross-validation.      

\begin{table}[!htb]
\caption{Performance Comparison of the DNN vs. Benchmarks ML Classifiers} 
\label{tab:cv10}
\begin{tabular}{lllllllll}
\toprule
 & P\&L   & AMC    & AUC   & Sensitivity & Specificity & Precision & G-Mean & F-score \\
 \hline
DNN                & \textbf{121.67} & \textbf{483.84} & \textbf{0.814} & \textbf{0.631}       & 0.994       & \textbf{0.987}     & \textbf{0.792}  & \textbf{0.661}   \\
Logit              & 111.67 & 926.45 & 0.705 & 0.293       & 0.997       & 0.982     & 0.540  & 0.402   \\
ANN                & 109.14 & 915.73 & 0.635 & 0.301       & 0.997       & 0.982     & 0.545  & 0.408   \\
Random Forest      & 107.18 & 928.61 & 0.720 & 0.291       & \textbf{0.997}       & 0.983     & 0.538  & 0.404   \\
AdaBoost           & 101.24 & 937.48 & 0.638 & 0.284       & 0.996       & 0.981     & 0.532  & 0.383   \\
SVM                & 68.34  & 825.56 & 0.691 & 0.370       & 0.995       & 0.984     & 0.598  & 0.430   \\
\bottomrule \bottomrule
\end{tabular}
\end{table}

Table \ref{tab:cv10} provides evidence for the superiority of the proposed DNN. It performs substantially better than ML benchmarks across several performance indicators. The only exception is specificity where all classifiers perform close to perfect and RF has a small edge over the other models. In terms of P\&L, the most relevant indicator for STX, we observe an improvement of nine percent compared to the logit model; the runner-up in terms of P\&L.  Improvements in the second monetary metric, AMC, are more substantial. The DNN performs roughly twice as good as ML benchmarks. 
Other indicators in Table \ref{tab:cv10} clarify the origin of the AMC results. All classifiers achieve high specificity and precision, while we observe sizeable differences in terms of sensitivity. Precision close to one indicates that whenever a model classifies a trade to pose high-risk and recommend hedging, this classification is typically correct. This is true for every classifier in the comparison. However, the DNN achieves higher sensitivity, meaning that it identifies a larger fraction of such high-risk trades. In consequence, the DNN hedges against more trades that come from A-book clients and avoids corresponding losses. In this regard, one may interpret Table \ref{tab:cv10} as evidence that the DNN implements a more prudent risk management strategy than the benchmarks. Results in terms of G-mean and the F-measure confirm that the DNN achieves a much better balance between sensitivity and specificity and precision, respectively. AUC results complement the evaluation and show that the DNN achieves better ranking performance. Ranking trades according to model-estimated class probabilities, the DNN discriminates A- and B-client trades more accurately than the benchmarks, achieving 13 percent higher AUC than the (second best) RF classifier. \\
In relation to monetary performance, it is interesting to consider why the DNN displays a larger edge over benchmarks in terms of AMC than in P\&L. The two measures differ by construction. P\&L  considers the actual return from a trade, whereas AMC is based on the average return over A-client and B-client trades, respectively. Considering a trade that leaves the market maker with a large loss, P\&L will penalize a wrong classification of that trade more severely than AMC. Therefore, Table \ref{tab:cv10} indicates that the DNN has misclassified some large loss trades, which benchmark models predicted correctly. This might suggest that an ensemble of the DNN and other classifiers could perform even better than the DNN alone. However, exploiting synergy among different classifiers is not trivial if their individual performances are very different. For example, it is clear from Table \ref{tab:cv10} that a simple average of the forecasts of the DNN and one other classifier will deteriorate predictive accuracy. Given the appealing performance of the DNN in Table \ref{tab:cv10} we leave an evaluation of forecast combination for future research and focus our attention on clarifying the origins of DNN performance.

The performance of a classifier depends on the features and their predictive value. The explanatory data analysis did not identify single features with high predictive power. A simple logit model using only the features of Table  \ref{tab:eda} performed close to random. To obtain a clearer view on the feature-response relationship, we employ information-fusion based sensitivity analysis (IFBSA). Originally proposed by \citep{ifbsa_first}, IFBSA has been used in several studies to examine feature importance through the lens of multiple classifiers \citep{oztekin_18,
ejor_sevim_14, jcis_oztekin_14}. IFBSA first assesses the importance of an individual feature as the percentage ratio of the model error without this feature to the model error with that feature included \citep{ejor_sevim_14}. IFBSA repeats this assessment for all features. Evaluating the marginal impact of a feature on model performance, the first IFBSA step shares similarities with RF-based feature importance weights, which interested readers can find in the online Appendix. A crucial distinction between IFBSA and RF-based feature importance is that the former considers several different classification models (i.e., the classifiers of Table \ref{tab:cv10} in our case). Let $s_t \left ( \boldsymbol{x} \right )$ denote the prediction of classifier $t=1, ..., T$ for feature vector $\boldsymbol{x}$ (e.g., $p \left ( \text{A-client} | \boldsymbol{x} \right )$). We can then write the combined prediction of a committee of $T$ classifiers as:
\begin{align}
\hat{y}_{fused}=\psi \left ( s_1 \left ( \boldsymbol x \right ), s_2 \left ( \boldsymbol x \right ), ..., s_T \left ( \boldsymbol x \right ) \right ) 
\label{ifbsa_fusion_fun}
\end{align}
where $\psi$ represents a mechanism to fuse individual classifier predictions. Considering a linear fusion function with normalized weights $\omega_t$ such that  $\sum_{t=1}^{T}\omega_t=1$, equation (\ref{ifbsa_fusion_fun}) becomes \citep{dss_Oztekin_13}:
\begin{align}
\hat{y}_{fused} = \sum_{t=1}^{T}\omega_t s_t\left ( \boldsymbol x \right )
\label{ifbsa_linreg}
\end{align}
IFBSA determines classifier weights $\omega_t$ based on the predictive performance of a classifier. This ensures that better classifiers such as the DNN in Table \ref{tab:cv10} receive higher weights while all classifiers' opinions on a feature are taken into account. More specifically, let $V_{t,k}$ denote the normalized feature importance weight of feature $k$ in classifier $t$. Then, replacing predictions $s_t\left ( \boldsymbol x \right)$ in (\ref{ifbsa_linreg}) with $V_{t,k}$, we obtain the final, fused importance weight of that feature as \citep{oztekin_18}:
\begin{align}
V_{k}\left ( fused \right ) = \sum_{t=1}^{T}\omega_t V_{t,k}\left ( \boldsymbol x \right )
\label{ifbsa_final}
\end{align}

\begin{figure}
\centering
\begin{minipage}[t]{0.50\columnwidth}
 \includegraphics[height=3in,width=0.9\textwidth]{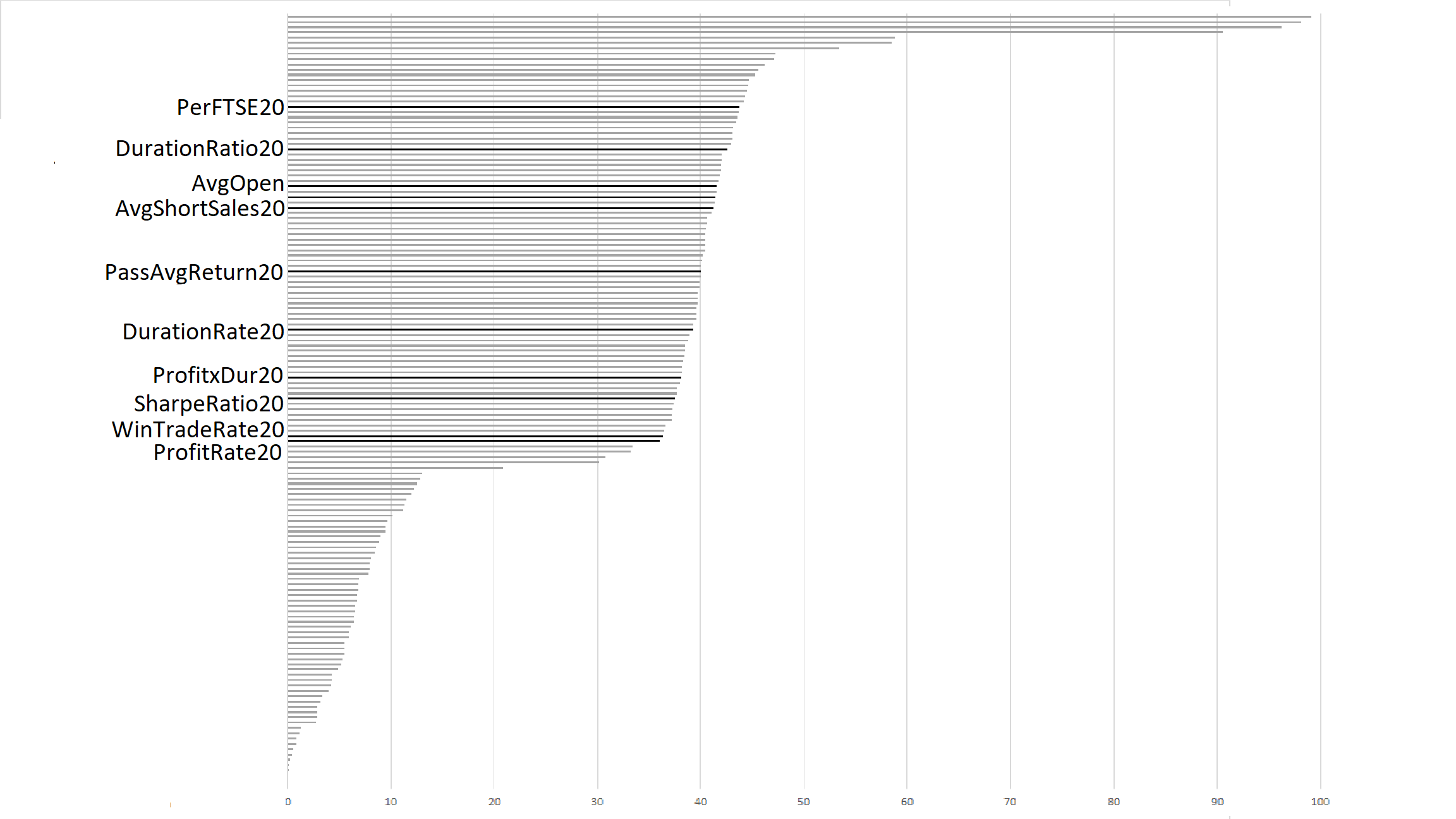}
\captionsetup{justification=centering}
\caption{Feature importance according to IFBSA \citep{ifbsa_first, ejor_Oztekin_16}}
\label{fig:ifbsa}
\end{minipage}
\begin{minipage}[t]{0.45\columnwidth}
 \centering
\includegraphics[height=1.75in,width=1\textwidth]{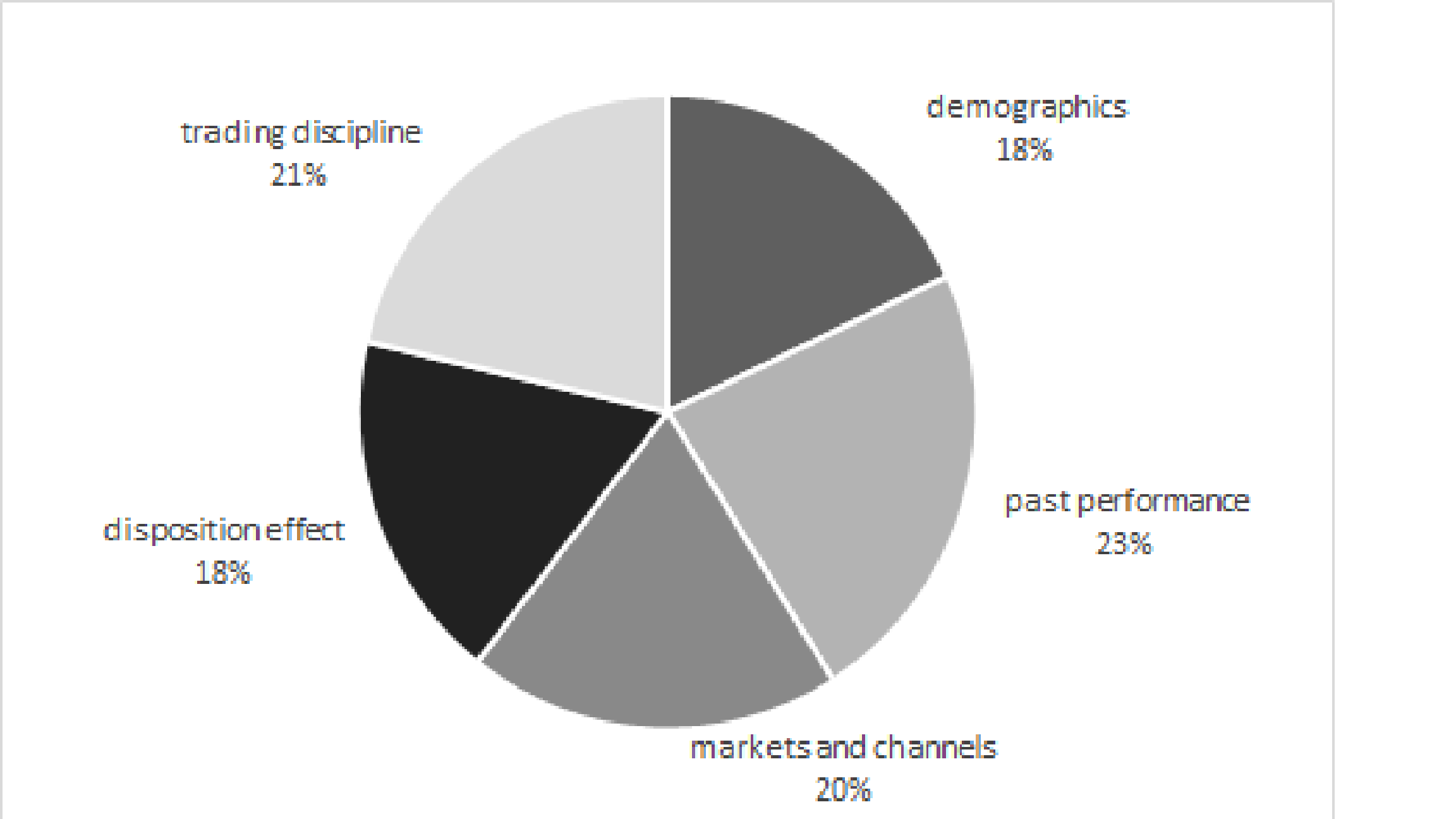}
 \captionsetup{justification=centering}
\caption{IFBSA importance per feature group.} \label{fig:ifbsa_group}
\end{minipage}
\end{figure}

Applying IFBSA to the classifier results of Table \ref{tab:cv10}, we obtain the distribution of fused feature importance weights shown in Figure \ref{fig:ifbsa}. We highlight features that the Fisher-score-based importance analysis (see Table \ref{tab:eda}) identified as the top ten most important features. Figure \ref{fig:ifbsa} depicts aggregated IFBSA importance weights for each of the five groups of features in our data set.

The distribution of feature importance in Figure \ref{fig:ifbsa} displays two break points. There is a small set of highly informative features. We also observe a large set covering roughly half of the remaining features where importance ranks are moderate. IFBSA attributes relatively low importance weights to the remaining features. The top-ten features of the Fisher-score assessment (see Table \ref{tab:eda}) belong to the second group. IFBSA judges these to be moderately important. Given that IFBSA and the Fisher-score incorporate fundamentally different perspectives toward feature importance, differences between their importance rankings are to be expected. We take Figure \ref{fig:ifbsa} as evidence of intricate dependencies between the binary response and the features, which the Fisher-score does not capture. For example, the Fisher-score ranking does not incorporate those features that IFBSA finds most relevant. It is tempting to characterize the three groups of features in Figure \ref{fig:ifbsa} as groups including  features that are, respectively, nonlinearly related to the prediction target, linearly related, and unrelated. However, Table \ref{tab:cv10} challenges this interpretation in that the linear logit model performs competitively to nonlinear classifiers. ANN, SVM, RF, etc. should be able to extract predictive value from the top-ranked features in Figure \ref{fig:ifbsa} beyond that which the logit model is able to extract if these features were nonlinearly related to the target. Yet, across several indicators of predictive performance, their forecasts do not appear to be more accurate than those of the logit model. 
Recall that IFBSA discounts feature importance signals of all classifiers but weights these according to classifier performance when computing final, fused feature importance weights via (\ref{ifbsa_linreg}) and (\ref{ifbsa_final}). Table \ref{tab:cv10} shows the DNN to perform substantially better than the benchmarks. Accordingly, fused weights are slightly geared toward DNN results ($\omega_{DNN}\approx 20$ percent). Therefore, we expect the ranks in Figure \ref{fig:ifbsa} to also reflect a feature's contribution to a latent, abstract representation of traders' behavioral characteristics, which the DNN distills from the data through its deep architecture. However, this interpretation requires further analysis to better understand the origin of DNN success and how it connects to the features. Subsequent sections shed light on these matters. To conclude the feature importance evaluation using IFBSA we note from Figure \ref{fig:ifbsa_group} that the overall amount of predictive information within the features distributes evenly across the five categories of features.
\color{black}
\subsection{Antecedents of DNN Forecast Accuracy}
Table \ref{tab:cv10} evidences the superiority of the DNN over ML-benchmarks for the specific data set used here. To examine the robustness of model performance, it is important to clarify the antecedents of DNN success. One characteristic of the DNN is its multilayered - deep - architecture. Prior research establishes a connection between the depth of a model and its expressive capacity \citep{montufar2014number}, which suggests depth to be a determinant of predictive accuracy. The use of unsupervised pre-training also distinguishes the DNN from the ML benchmarks. Aiding model training through finding more abstract, generative features, we expect predictive accuracy to benefit from pre-training. A third factor of interest is class imbalance. Skewed class distributions are a well-known impediment to classification. The DNN being more robust toward class imbalance could thus also explain the results of Table \ref{tab:cv10}. In the following, we examine the influence and importance of these three factors.

\subsubsection{The Deep Architecture}
The DNN generates predictions in the last layer, where the last layer output neuron receives the combined input from multiple previous layers of SdAs and translates these signals into class probability predictions using the softmax function. This network configuration is equivalent to running logistic regression on the output of the hidden layers. To shed light on the value of the deep architecture, we compare DNN predictions to predictions from an ordinary logistic regression with the original features as covariates. The logistic model represents an approach which takes away the deep hidden layers from the DNN and only sustains the last layer. This is useful for appraising the merit of the distributed representations, which the deep hidden layers extract from the raw features.

\begin{figure}[H]
\begin{center}
\includegraphics[height=3in,width=0.5\textwidth]{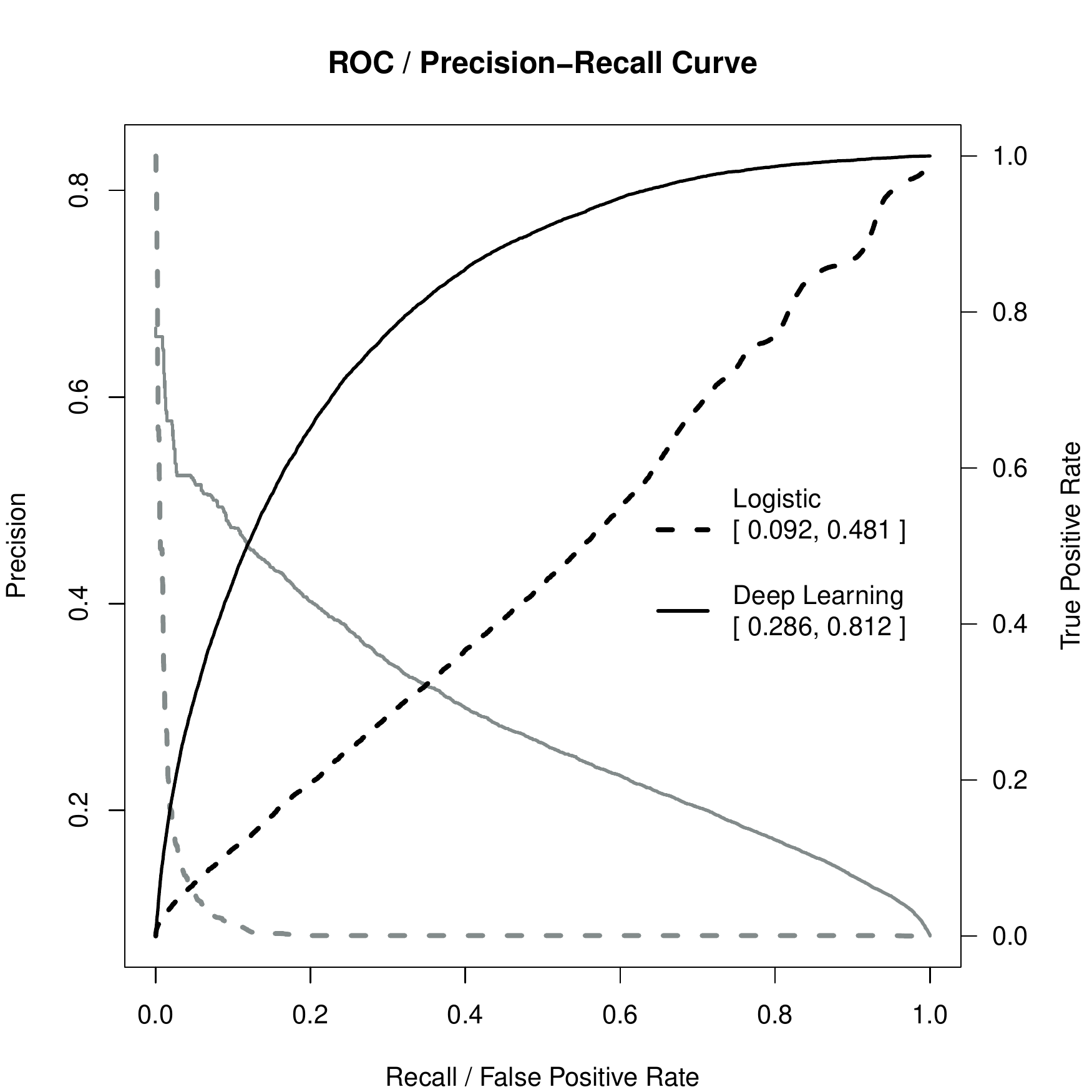}
\caption{ROC (black), Precision-Recall Curve (grey)  of DL  and logit.} \label{logis}
\end{center}
\end{figure}

Figure \ref{logis} displays the receiver-operating-characteristics (ROC) and a Precision-Recall (PR) curve for the DNN and a simple logit model. The plot emphasizes that the deep architecture substantially improves the network's discriminative ability. The performance of the logit model on raw features is almost random. The AUC value of 0.812 for the DNN suggests that performing the same regression on the high level representations, which the DNN learns from the raw features, facilitates a reliable detection of the positive class. Consequently, the DNN succeeds in extracting predictive features from the  input data. In appraising Figure \ref{logis} it is important to note that the logit model is not meant to contribute a strong benchmark. As shown in Table \ref{tab:cv10}, a regularized logit model with feature selection performs better than random. The purpose of Figure \ref{logis} is to evaluate the overall effect of the deep architecture compared to using the raw features as is, which motivates using the ordinary logit model for this comparison. The overall conclusion emerging from the analysis is that the deep architecture affects the predictive performance of the DNN.

\subsubsection{Unsupervised Pre-Training}
\begin{figure}
\centering
\begin{minipage}[t]{0.50\columnwidth}
 \includegraphics[width=0.9\textwidth]{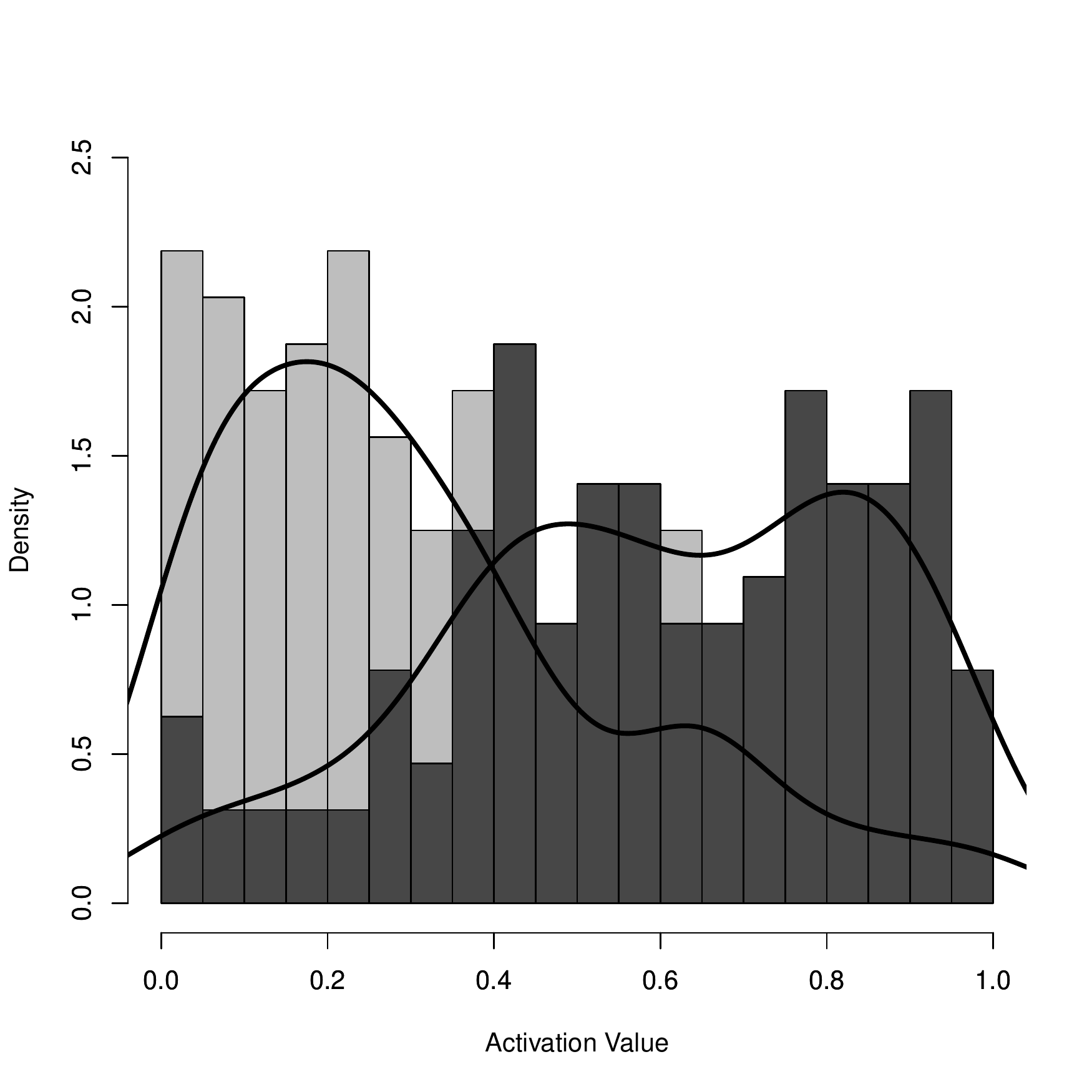}
\captionsetup{justification=centering}
\caption{Histogram of first dA layer activation values for A-book (dark color) and B-Book (light color) client trades.} 
\label{activation}
\end{minipage}%
\begin{minipage}[t]{0.40\columnwidth}
 \centering
\includegraphics[width=1\textwidth]{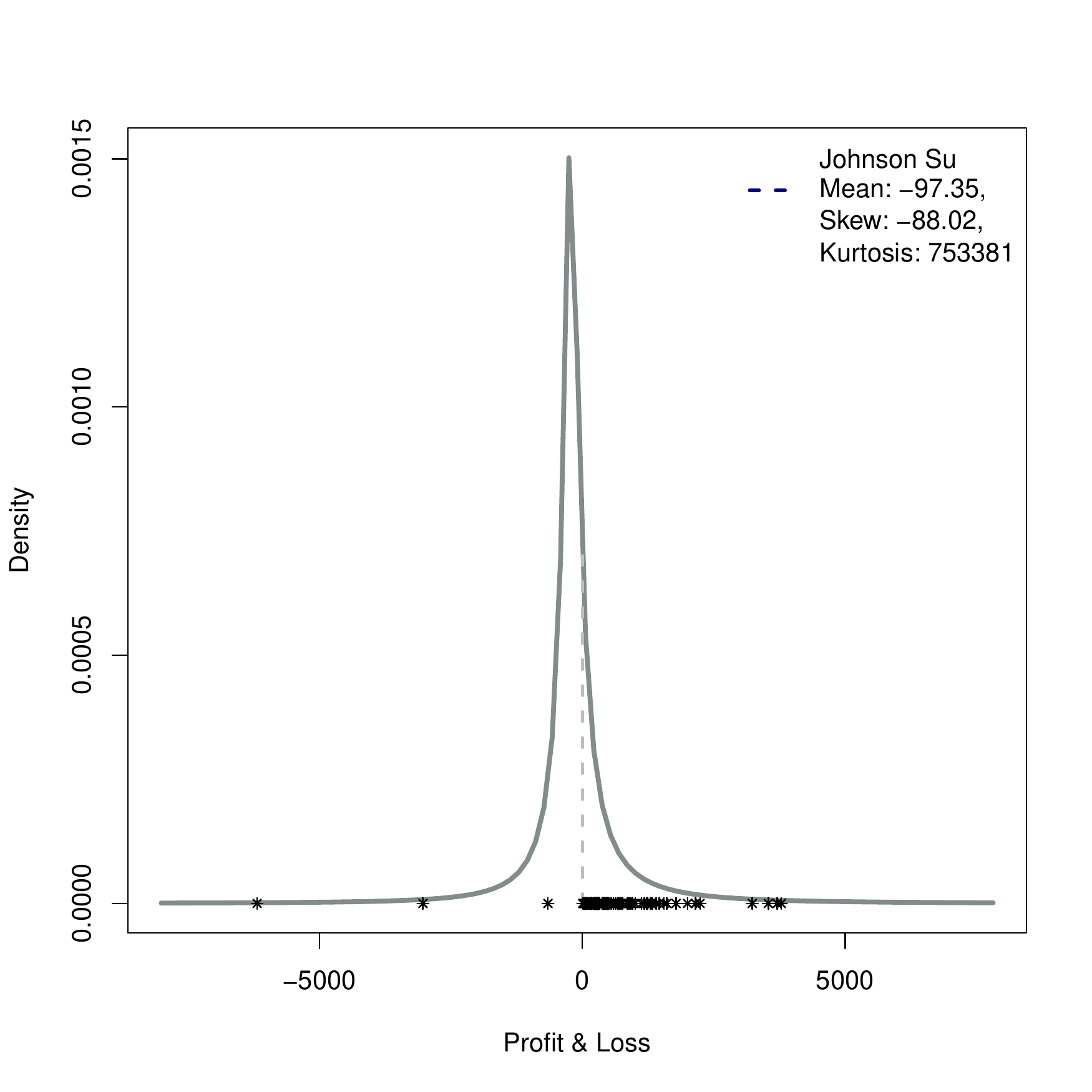}
 \captionsetup{justification=centering}
\caption{Top $100$ stimuli of the best neuron from the test set } \label{stimulus}
\end{minipage}
\end{figure}

The proposed DNN uses unsupervised pre-training for representation learning and feature extraction. To confirm the merit of pre-training, we examine the discriminative strength of each neuron in the unsupervised pre-training stage. We aim to check whether DNN learns distributed representations that help differentiate A- and B-book clients from \emph{unlabeled} data. To that end, Figure \ref{activation} provides the histograms of activation values for neurons in the first dA layer of the DNN. The histograms show that activation values tend to be less than $0.4$ when receiving a trade from a B-book client. Trades from A-book clients typically result in an activation value of $0.4$ and above. While the magnitude of the activation values is irrelevant, the discrepancy of activation values for trades from different types of clients illustrates that - even with unlabeled data - the neurons in the first dA layer differentiate A- from B-book client trades. The intricate non-linear transformation between layers prohibit a replication of this analysis for higher layers because the relationship between activation values and input signals is no longer monotone. However, Figure \ref{activation} provides preliminary evidence that the spread trading data facilitates the extraction of higher-level generative features using pre-training.


To gain more insight into the link between activation values and trades from different types of traders, we examine whether trades that trigger high activation values in a neuron are worth hedging. We first calculate the maximum and minimum activation values for every neuron of the first layer, and $20$ equally spaced threshold values between these boundaries. Subsequent analysis is based on a single neuron. We chose the neuron and corresponding threshold that give the purest separation between A- and B-book client trades (see Figure \ref{activation}) upon manual inspection. Using this neuron, we find the $100$ trades that activate the neuron the most. Figure \ref{stimulus} plots these trades on the overall P\&L distribution. The results illustrate that, with a few false negatives, $97\%$ of the trades that maximally activate the neuron end in profit and leave the market maker with a loss. Hedging against these trades, as indicated by the neuron's activation levels, is economically sensible. Although the eventual hedging decisions are based on the prediction of the DNN as a whole, the single neuron analysis provides further evidence of unsupervised pre-training of SdA layers to extract patterns that are indicative of a trade's risk. This confirms that the DNN learns distributed representations from the input data, which eventually help to distinguish high risk traders from other clients.


\subsubsection{Analysis of the Class Imbalance Effect}
\label{sec:imbalance}
A growing body of literature on deep imbalanced learning indicates that DL models inherit vulnerability toward class imbalance from their ML ancestors \citep{dl_imbalance_survey}. On the other hand, pre-training might equip the proposed DNN with some robustness toward class skew. Pre-training is carried out in an unsupervised manner so that class imbalance cannot impede model training. Figure \ref{activation} indicates that, without having access to class labels, pre-training has extracted patterns that help to differentiate high risk traders and B-book clients. Only the DNN has access to this information, which might give it an advantage over the ML benchmarks in the comparison of Table \ref{tab:cv10}. To test this, we repeat the comparison after addressing class imbalance using the SMOTE (synthetic minority class oversampling technique) algorithm. SMOTE remedies class skew through creating artificial minority class examples in the neighborhood of actual minority class cases \citep{he2009learning}. We execute SMOTE until the number of artificially created A-book clients equals that of the B-book clients, and repeat the comparison of classifiers on the oversampled data. Since SMOTE increases the number of observations and, in turn, the time to train a model, we reduce the number of cross-validation folds to five. Note that oversampling occurs only in the training data. The validation fold reflects the true class distribution in the data to give an unbiased view on model performance. Table \ref{tab:smote} details the classifiers' predictive performance on the oversampled data.

\begin{table}[]
\caption{Predictive Performance of the DNN vs. Benchmark ML Classifiers After Applying SMOTE} 
\label{tab:smote}
\begin{tabular}{lllllllll}
\toprule
 & P\&L   & AMC    & AUC   & Sensitivity & Specificity & Precision & G-Mean & F-score \\
\hline
DNN           & \textbf{283.31} & \textbf{243.17} & \textbf{0.889} & \textbf{0.815}       & 0.996       & \textbf{0.992}     & \textbf{0.896}  & \textbf{0.804}   \\
Logit         & 260.45 & 525.57 & 0.806 & 0.600       & 0.983       & 0.979     & 0.751  & 0.397   \\
ANN           & 233.55 & 484.19 & 0.858 & 0.632       & 0.974       & 0.967     & 0.785  & 0.438   \\
Random Forest & 262.85 & 581.22 & 0.792 & 0.557       & 0.988       & 0.979     & 0.742  & 0.519   \\
AdaBoost      & 259.45 & 484.19 & 0.803 & 0.632       & 0.974       & 0.967     & 0.785  & 0.438   \\
SVM           & 204.35 & 930.26 & 0.825 & 0.619       & \textbf{0.992}       & 0.984     & 0.779  & 0.522   \\
\bottomrule \bottomrule
\end{tabular}
\end{table}

 We find SMOTE to improve all classifiers. With very few exceptions, results in Table \ref{tab:smote} display higher performance compared to Table \ref{tab:cv10} and the magnitude of the improvement is often substantial. For example, the AUC of the DNN is nine percent higher (0.889 c.f. 0.814) when using SMOTE. Other classifiers experience a similar or larger improvement. The ANN classifier, for example, achieves an AUC that is 35 percent better with SMOTE (0.858 c.f. 0.635). The sensitivity results are substantially better, which confirms that SMOTE has increased a classifier's awareness of the minority class. Higher sensitivity often comes at the price of reduced specificity and precision. However, Table \ref{tab:smote} reveals that classifier performance in terms of both specificity and precision is almost as good as before applying SMOTE. Accordingly, the G-mean and F-score emphasize that SMOTE achieves a much better balance between sensitivity (recall) and specificity and precision, respectively. Likewise, P\&L and AMC are magnitudes better with SMOTE.

We also find some evidence that the margin by which the DNN outperforms ML benchmarks decreases. Considering for example the AUC, which gives an overall assessment of a classifiers discriminatory ability, the strongest benchmarks after oversampling is ANN, which the DNN outperforms by a margin of 3.6 percent (0.889 c.f. 0.858). In Table \ref{tab:cv10}, the edge of the DNN over the runner-up in AUC was 13 percent (0.814 c.f. 0.720 for RF). 
An interpretation of this tendency is that Table \ref{tab:cv10} gives an optimistic picture of DNN performance. The accuracy gap between the DNN and the ML benchmarks is less than that which Table \ref{tab:cv10} suggests if ML benchmarks receive auxiliary tuning in the form of remedying class imbalance using SMOTE. However, this view also implies that the DNN to be more robust toward class imbalance. While benefiting from SMOTE, its ability to identify high risk traders accurately is less dependent on oversampling the minority class compared to the ML benchmarks. \color{black}This interpretation follows from examining the magnitude of performance improvement in Table \ref{tab:smote} compared to Table \ref{tab:cv10}, which is typically smaller for the DNN than for ML benchmarks. Higher robustness toward imbalance agrees with results of Figure \ref{activation} concerning the merit of unsupervised pre-training.

\subsection{Implications for Risk Management}
A model-based hedging policy comprises hedging the trades of clients classified as A-book by the model and taking the risk of all other trades. To clarify the managerial value of the proposed DNN, we compare its P\&L to that of rule-based hedging strategies. One rule-based approach is the current policy of STX, which involves hedging trades of clients who secured a return above five percent in their previous 20 trades. In addition, we develop three custom hedging heuristics. Our first policy, \emph{Custom 1}, relies on the Sharpe Ratio and 
singles out traders who achieve a higher than average Sharpe ratio in their past 20 trades. We suggest that securing risk-adjusted returns above the average indicates trader expertise. Since professionalism is only one reason for a successful trading history, \emph{Custom 2} heuristic addresses another group of traders, which we characterize as overconfident. Such traders may display higher yields than other market participants and exhibit aggressive trading behavior, manifesting itself through bigger lot sizes, higher frequency and shorter time interval trades \citep{BENOS1998353}. The Custom 2 heuristic thus considers the average trade duration and number of trades to deduce traders who may pose a greater risk. The third strategy, \emph{Custom 3}, hedges trades from clients with a positive track record since trading with STX. The rationale is that traders who are unsuccessful in their early experiences might quit. Traders with a longer track record are either truly successful (and should be hedged against) or gamblers with a negative expected value (and should not be hedged against). Following this line of thinking, the most important risk STX is facing comes from \emph{new} A-book clients. Comparisons to Custom 3 shed light on the ability of the DNN to identify such new A clients, as improvement over Custom 3 signals the DNN recognizing high risk traders that the track record-based logic of Custom 3 fails to capture.\footnote{We are grateful to an anonymous reviewer who suggested the logic of the Custom 3 heuristic.} We also consider an ensemble of the custom rule-based heuristics, constructed by means of majority voting. 

Drawing on domain knowledge, the rule-based strategies adopt a deductive approach. To complement the analysis, we add an inductive approach in the form of a classification tree. Trees are regarded as interpretable classifiers. However, the degree to which decision makers can understand trees depends on their depth. In the interest of interpretability, we consider a classification tree (ctree) with two levels. 

\begin{figure}
\includegraphics[height=2.5in,width=0.9\textwidth]{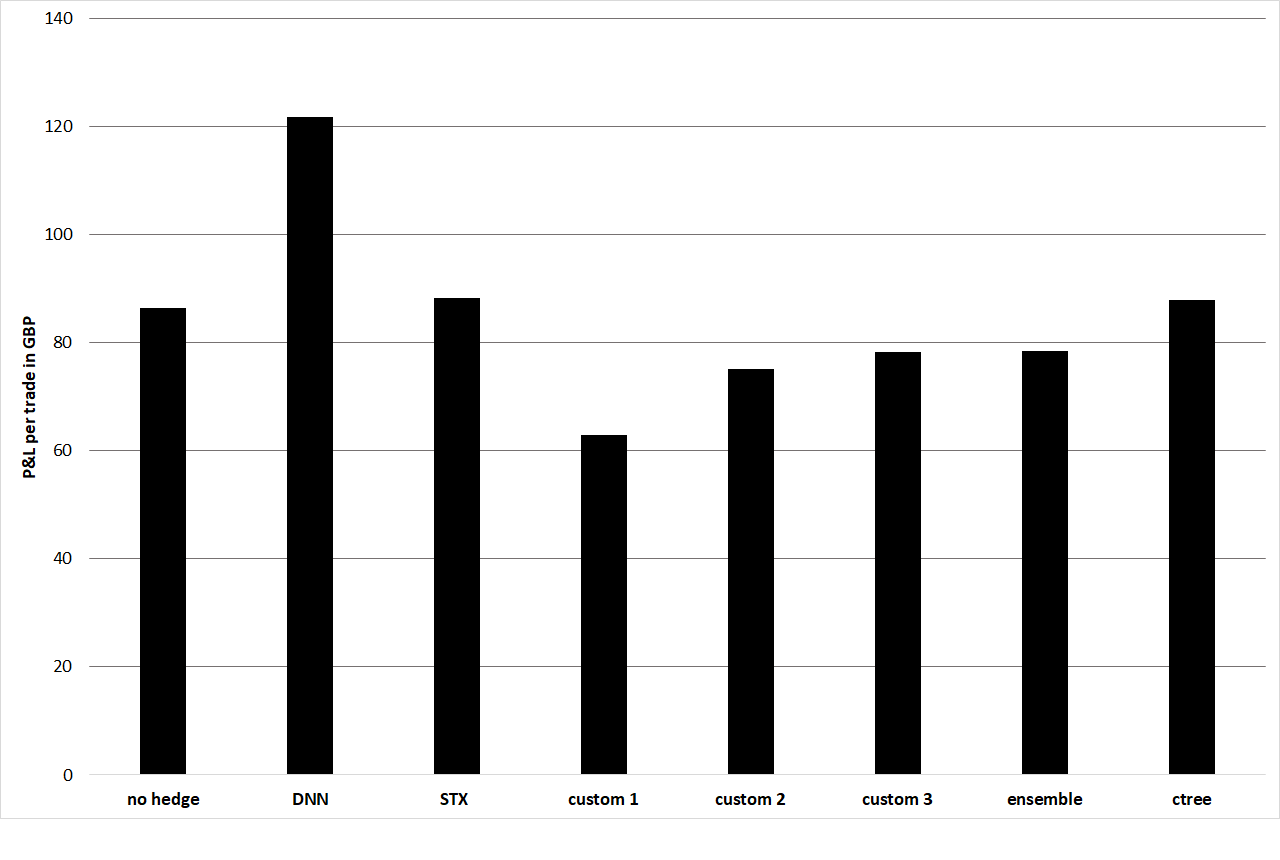}
\caption{Average P\&L per trade in GBP of the DNN and Rule-Based Hedging Policies} 
\label{fig:PnL_heuristics}
\end{figure}

Figure \ref{fig:PnL_heuristics} reveals the current policy of STX to be the most suitable deductive hedging strategy. The logic of Custom 1 - 3 draws on financial theory. However, each of the three approaches, as well as an ensemble of them, performs worse than a hypothetical baseline setting in which STX would not hedge any trade. Observing Custom 1 - 3 to perform  worse than this baseline supports the view that the focal trader classification task represents a challenging problem. Following this line of reasoning, Figure \ref{fig:PnL_heuristics} also emphasizes the soundness of the STX policy. 

Unlike the deductive STX approach, the tree-based heuristic learns from past data. We observe the tree to use three features: a trader's average P\&L in their initial 20 trades with STX, the minimum number of minutes until closing a losing position in their last 20 trades, and the average Sharpe ratio over their last 20 trades. These features display similarity with the custom heuristics. For example, considering a trader's initial performance follows the logic of Custom 3. Accounting for risk-adjusted returns is similar to Custom 1. Finally, considering a trader's reaction toward losses, the tree uses one of the variables capturing the disposition effect. We observe the two-level tree to produce slightly larger P\&L than the STX heuristic. This suggest that a trader's average past performance, embodied in the STX approach, approximates the more complex rule set of the tree with some accuracy. Although the criticality of accurate hedging in the spread trading market suggests a revision of the STX approach with a tree-based approach, another finding from Figure \ref{fig:PnL_heuristics} is that implementing a DNN-based hedging strategy enables STX to further improve P\&L compared to its current policy and the other rule-based hedging strategies we consider. Compared to the STX heuristic,  the DNN raises per trade profits by 121.67 - 88.28 = 33.40 GBP, which implies a substantial improvement.  \color{black} 

The STX heuristic represents an established business practice at the partner company and reflects many years of industry experience. Moreover, the heuristic is extremely fast to execute and completely transparent. The situation for the DNN is far different. Classifying incoming trades more accurately, a DNN-based hedging policy is more profitable than rule-based approaches. The main cost of accuracy and profitability improvements is the black-box character of the corresponding risk management system. The client classification rules from the DNN originate from automatically extracted distributed representations of high risk traders. The business logic encapsulated in these rules is not interpretable for decision-makers, which also prohibits testing the agreement of these rules with domain knowledge. 

Improved performance of the DNN leaves risk managers with the task to decide whether performance improvements are large enough to compensate the opaqueness of DNN and associated disadvantages, such as a lack of justifiability, higher computational requirements, etc. In the case of STX, we expect the imperative to hedge trades accurately and the magnitude of the performance improvement observed on their data to justify the adoption of a sophisticated DNN-based hedging strategy. The same might be true for other the spread-trading companies, although these would first need to replicate the results of this study to confirm the effectiveness of the DNN. A detailed description of the DNN configuration in the paper and especially the online Appendix will simplify this task  (see Algorithm 1 in the online Appendix)\color{black}. A more strategic consideration is that reluctance to adopt a new technology such as a sophisticated DNN-based hedging policy might harm the competitive position of STX if competitors deploy corresponding solutions and use them to offer better prices to retail investors. At the same time, we caution against an overly optimistic view toward advanced DL-based decision aids. The empirical results observed in this study come from a single data source, which, although large in size, reflects the peculiarities of the market position and client structure of STX, and require a replication with different data in future research to raise confidence in the superiority of DL that we observe here.

\section{Discussion}
\begin{figure}
\centering
\begin{subfigure}{.45\textwidth}
  \flushleft
  \includegraphics[height=1.7in]{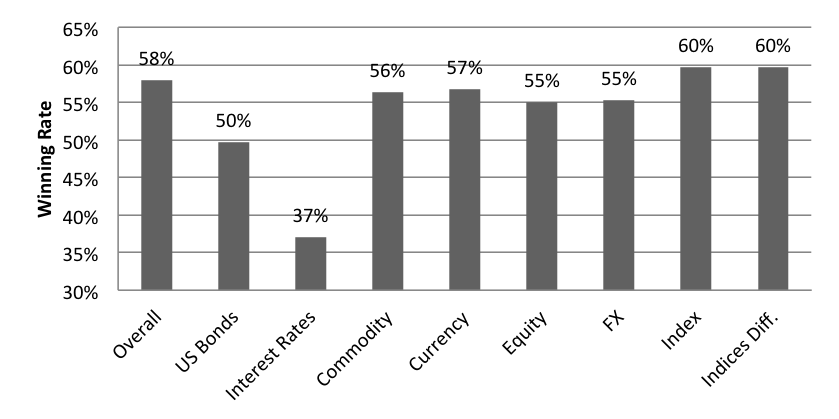}
\end{subfigure}
\begin{subfigure}{.45\textwidth}
  \flushleft
  \includegraphics[height=1.7in]{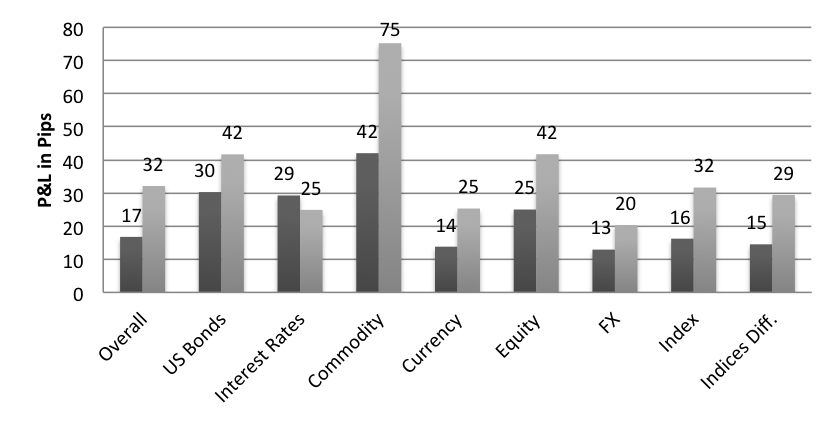}
\end{subfigure}
\caption{Retail traders' average winning ratio and average P\&L points (profit in dark, loss in grey) on different categories of investments on the spread trading market. }
\label{plstat}
\end{figure}

The empirical results suggest the DNN to outperform rule-based and ML benchmarks. It identifies high risk traders more accurately than other classifiers and provides higher financial gains when used for hedging decisions. 
Striving to explain the inferior performance of ML benchmarks, we note that predicting traders' risk taking behavior and future profitability under dynamic market conditions is a challenging task. Traders differ in their characteristics and trading behavior, and both are likely to change over time. Identifying unskilled traders is especially difficult due to the high variation in both behavior (input) and performance (output). 
Compared to genuine good traders, it is harder to identify uniform trading patterns for poor traders. Interviews with STX's staff hint at skilled traders sharing certain characteristics such as the ability to capture market rallies, following a consistent strategy, setting and adjusting limits, etc.
There are countless ways in which poor traders lose money, including ignoring any of the above rules. In the high dimensional behavioral space, the number of potential variations of poor traders is innumerable. This contradicts the prior assumption of ML methods that the distribution $P(label | features)$ is smooth and well represented in the training data. Although based on an economic rationale, input features relating to past risk-adjusted return, trading frequency, etc. do not facilitate an accurate discrimination of spurious from genuine good traders. This arises because several feature values may coincide. The entanglement of spurious and genuine good traders in the behavioral feature space of trader characteristics complicates the trader classification problem. Moreover, the chance of making profit in the spread-trading market is highly noisy. Even poor traders can, by luck, win money. In fact, Figure \ref{plstat} reveals that most of the clients who trade with STX have a greater than fifty percent win/lose ratio. However, Figure \ref{plstat} also shows that average losses exceed average winnings by a large margin. Therefore, it is often sensible to classify a trader as a B-client and refrain from hedging their trades, even if many of their previous trades ended in profit. 

The deep architecture equips DNN with higher expressive capacity to store the large number of variations of trading behaviors. Increased depth enables the DNN to profile new combinations of behavioral variations and generalize to new trading patterns less represented in the training data. Furthermore, a specific DNN component we employ for trader classification is unsupervised pre-training. Observed results confirm that pre-training enables the DNN to construct layers of feature detectors that capture underlying generative factors, which explain variations across different trading behaviors. Stacking multiple layers of progressively more sophisticated feature detectors, the DNN learns to disentangle these factors from the input distribution. Variations that are important for subsequent discrimination are amplified, while irrelevant information within the input data is suppressed \citep{erhan2010does}. We examine this ability in Figure \ref{logis}, \ref{activation} and  \ref{stimulus}. After pre-training, the higher levels of the feature hierarchy store robust, informative, and generalizable representations that are less likely to be misled - and, thus, invariant to - the entangling of trading patterns in the input-space. In summary, the DNN draws upon the raw features and creates sensible abstractions from these features that exhibit a stronger connection with the target and facilitate predicting that target with higher accuracy.

\section{Conclusions}
We set out to examine the effectiveness of DL in management support. Corresponding applications often involve developing normative decision models from structured data. We focus on financial risk taking behavior prediction and develop a DNN-based risk management system.

The results obtained throughout several experiments confirm the ability of DL, and the specific architecture of the DNN we propose, to extract informative features in an automatic manner. We also observe DNN-based predictions of trader behavior based on these features to be substantially more accurate than the forecasts of benchmark classifiers. Finally, our results demonstrate that improvements in forecast accuracy translate into sizable increases in operating profit. This confirms the ability of the proposed DNN to effectively support (hedging) decision-making in our risk management case study.

More generally, we provide evidence that the characteristic features of DL may generalize to the structured data sets commonly employed in retail finance and corporate decision support. In particular, we demonstrate that it is possible to successfully model patterns associated with individual behaviour (which can be variable, erratic and dynamic), in an automatic manner to generate features that determine a target measure (e.g. profitability). For the case study explored here, we faced the challenges of class imbalance, concept drift (which arises in dynamic environments), the curse-of-dimensionality and the high costs to develop and revise hand-crafted features in response to external changes.  Many of these challenges are faced by those seeking to predict human behaviour in related fields. Until now, most of these tasks have been tackled using a range of statistical and ML techniques. Our findings, open up the possibility of employing a similar DL architecture to that adopted here to tackle a range of consumer related prediction tasks, such as helping banks improve their risk control by making effective credit approval decisions and predicting the solvency of corporate borrowers, of assisting governments predict how consumers will respond to tax and regulation changes, and supporting companies in marketing decisions involving churn prediction, response modelling and cross-/up selling. Many of these tasks rely on structured data, and we have shown that the DL methodology we propose can improve forecasting accuracy in settings that require the prediction of human behaviour based on structured data. 

A distinctive feature of our case study stems from the fact that a small group of individuals pose very high risk. In theory, a single trade can result in unbounded losses in the spread trading market. High risk individuals (A-clients) are only a small fraction of the population but determine the overall risk exposure of the market maker to a large extent. This characteristic sets our study apart from standard credit scoring settings and other financial applications where losses are often bounded. Thus, our study contributes to new ways of managing major risks coming from minority individuals. DNNs may be particularly suitable for developing decision support system to tackle challenges with similar features such as insider or malicious behaviour detection, fraud modeling, or cyber-attack, where only a small set of professionals can cause major losses for an enterprise.  

A fruitful area for future research will be to explore to what extent the architecture adopted here proves effective and/or to what extent it needs to be adapted to meet the particular demands of other human behaviour-related forecasting tasks. Future work exploring the extent to which networks with supervised and unsupervised layers such as SdA, which we find to work well in risk analytics, is of particular interest to us since unsupervised learning offers advantages when labeled data is sparse. Having demonstrated that significant gains can be made in forecasting accuracy in one setting involving the prediction of human behaviour, our hope is that this leads to future studies demonstrating that DL can help solve many of the world’s problems associated with ‘unpredictable’ human behaviour.

\color{black}



\section*{Acknowledgement}
We thank the editor, Prof. Teunter, for his efforts in handling our paper and are  thankful to three anonymous reviewers whose feedback has helped tremendously to improve earlier versions of the paper. We are especially grateful to J.C. Moreno Paredes for his invaluable help with data preparation. 


\section*{References}
\bibliography{reference} 

\clearpage

\end{document}